\documentclass[%
 reprint,
preprintnumbers,
 amsmath,amssymb,
 aps,
pra,
]{revtex4-1}

\usepackage{lipsum}
\usepackage{graphicx}% Include figure files
\usepackage{dcolumn}% Align table columns on decimal point
\usepackage{bm}% bold math
\usepackage[dvipsnames,usenames]{xcolor}
\usepackage{amsmath}
\usepackage{amssymb} % AMS
\usepackage{hyperref}
\usepackage[qm,braket]{qcircuit}
\usepackage{bbold}
\usepackage{algpseudocode}
\usepackage{mathtools}

\begin{document}

\title{Short-depth circuits for efficient expectation value estimation}

\author{A. Roggero}
\email{roggero@uw.edu}
\affiliation{Institute for Nuclear Theory, University of Washington, 
Seattle, WA 98195, USA}

\author{A. Baroni}
\email{abaro008@odu.edu}
\affiliation{Department of Physics and Astronomy University of South Carolina,
712 Main Street, Columbia, South Carolina 29208, USA}

\preprint{INT-PUB-19-016}

\date{\today}% It is always \today, today,

\begin{abstract}
The evaluation of expectation values $Tr\left[\rho O\right]$ for some pure state $\rho$ and Hermitian operator $O$ is of central importance in a variety of quantum algorithms. Near optimal techniques developed in the past require a number of measurements $N$ approaching the Heisenberg limit $N=\mathcal{O}\left(1/\epsilon\right)$ as a function of target accuracy $\epsilon$. The use of Quantum Phase Estimation requires however long circuit depths $C=\mathcal{O}\left(1/\epsilon\right)$ making their implementation difficult on near term noisy devices. The more direct strategy of Operator Averaging is usually preferred as it can be performed using $N=\mathcal{O}\left(1/\epsilon^2\right)$ measurements and no additional gates besides those needed for the state preparation.

In this work we use a simple but realistic model to describe the bound state of a neutron and a proton (the deuteron) and show that the latter strategy can require an overly large number of measurement in order to achieve a reasonably small relative target accuracy $\epsilon_r$. We propose to overcome this problem using a single step of QPE and classical post-processing. This approach leads to a circuit depth $C=\mathcal{O}\left(\epsilon^\mu\right)$ (with $\mu\geq0$) and to a number of measurements $N=\mathcal{O}\left(1/\epsilon^{2+\nu}\right)$ for $0<\nu\leq1$. We provide detailed descriptions of two implementations of our strategy for $\nu=1$ and $\nu\approx0.5$ and derive appropriate conditions that a particular problem instance has to satisfy in order for our method to provide an advantage.
\end{abstract}


\maketitle
As we are approaching the era of noisy intermediate scale quantum devices (NISQ~\cite{Preskill2018}) the growing interest in practical applications of quantum computing techniques has led to an increased interest in algorithms that are robust against errors and require a limited number of qubits and gates.  A central component of many quantum computing algorithms, in particular for applications to quantum simulation~\cite{Feynman1982,Lloyd96}, is the estimation with error $\epsilon$ of the expected value of an hermitian operator $O$ on some quantum state described by the density matrix $\rho$ of a system of $n$ qubits:
\begin{equation}
\label{eq:exp_value}
\langle O\rangle = Tr \left[O\rho\right] = \bra{\Psi}O\ket{\Psi}\, ,
\end{equation}
where here and in the following we assume the qubits to be in a pure state described by $\rho=\ket{\Psi}\bra{\Psi}$. 
A common example is the calculation of the expectation value of the Hamiltonian of a many-body system which can be used, together with the variational principle, to guide the preparation of low energy states in algorithms such as the Variational Quantum Eigensolver (VQE)~\cite{Peruzzo2014,McClean2016,Kandala2017,Shen2017}. 
For unitary operators we can evaluate the expectation value in Eq.~\eqref{eq:exp_value} by simply measuring the state of the device in the appropriate basis. If the change of basis is expensive one can employ the Hadamard Test algorithm from~\cite{Ortiz01} which only requires a single application of $O$ conditional on an ancilla prepared in a superposition state.

In general however we are interested in evaluating $\langle O\rangle$ for hermitian operators $O$ (such as the Hamiltonian of many-body system) and thus alternative strategies have to be devised. Optimal quantum algorithms have been discovered in the past (see eg.~\cite{Knill07}), with efficiencies approaching the Heisenberg limit $N=\mathcal{O}(1/\epsilon)$ for the number of times the experiment needs to be repeated in order to achieve some target additive error $\epsilon$. This remarkable result is achieved by making effective use of both the Quantum Phase Estimation~\cite{Cleve98,Abrams1999} and Amplitude Amplification~\cite{Brassard2002} algorithms, along with the observation that we can estimate Eq.~\eqref{eq:exp_value} from the linear part of
\begin{equation}
\label{eq:smallt_exp}
\begin{split}
Tr\left[e^{i\tau O}\rho\right] &= \langle cos\left(\tau O\right)\rangle + i \langle sin\left(\tau O\right)\rangle\\
&= 1 + i\tau \langle O\rangle + \cdots
\end{split}
\end{equation}
up to additive errors that vanish in the limit $\tau\to0$.

The scheme requires $\mathcal{O}(1/\epsilon)$ applications of a controlled version of the unitary operator $U_\tau = e^{i\tau O}$ and it's inverse (for the QPE part) and $\mathcal{O}(1/\epsilon)$ application of the state preparation unitary $W$ (ie. $W\ket{0}=\ket{\Psi}$). Near term quantum devices will be characterized by a substantial level of noise~\cite{Preskill2018} and this will in general prevent us from employing these algorithms even in situations where performing $U_\tau$ is simple (eg. for operators $O$ diagonal in the computational basis) as the need to apply $W$ multiple times will lead to an excessively large gate count.

In general, schemes that are robust to noise are therefore preferable for early applications and this has led to a proliferation of hybrid quantum-classical algorithms~\cite{Bravyi2016} for a variety of purposes ranging from quantum simulation~\cite{Li2017}, to approximate optimization~\cite{Farhi2014} and quantum compiling~\cite{Khatri2019}. Our work follows the same philosophy in that we will delegate a substantial computational effort to classical computing resources while at the same time leveraging the available capabilities of current quantum hardware.

The paper is organized as follows. In Sec.~\ref{sec:SM} we review the standard method commonly used to evaluate expectation values of Hermitian operators known as Operator Averaging~\cite{McClean2014} and in Sec.~\ref{sec:newmethod} we present a detailed discussion of our proposed method based on a single step of phase estimation for performing this task. We then proceed in Sec.~\ref{sec:deuteron} to apply our approach to a simple but challenging nuclear physics problem: the calculation of the deuteron binding energy with a realistic interaction. We have devoted Sec.~\ref{sec:noise} to a more thorough exploration of the effect of noise on both methods and discuss more in depth why an ancilla-based scheme could provide potential benefits on near term noisy devices. As our method is not expected to be competitive in the asymptotically small error limit, in Sec.~\ref{sec:imp_ch} we describe how to overcome some of the major implementation challenges in assessing if our scheme could prove advantageous for a particular problem instance.
Finally in Sec.~\ref{sec:conc} we summarize the results and possible future directions of the present work. 

\section{Operator Averaging}
\label{sec:SM}
We discuss here the standard method commonly used in the literature to evaluate expectation values of Hermitian operators~\cite{Peruzzo2014,McClean2016} and start setting up the notations we will use throughout this work.
As a first step, and without loss of generality, let us separate out the traceless component $O_T$ from the observable
\begin{equation}
O = O_I + O_T\quad O_I \equiv \alpha_0 \mathbb{1}\quad Tr[O_T]=0
\end{equation}
for some $\alpha_0\in \mathbb{R}$. The idea behind the Operator Averaging (OA) approach first proposed in~\cite{Peruzzo2014} is to exploit a decomposition of $O_T$ of the following form
\begin{equation}
\label{eq:op_exp}
O_T = \sum_{k=1}^L \alpha_k U_k = \sum_{k=1}^L \alpha_k \left[e^{i\theta_k}P_k\right]\quad,\;\; a_k>0\;\forall k
\end{equation}
where $P_k\in\{\mathbb{1},X,Y,Z\}^{\otimes n}$ are tensor products of Pauli matrices and the angles $\theta_k$ are introduced in order to keep the coefficient $\alpha_k$ positive. The expectation value of each of the terms in the expansion can be estimated with $O(1)$ circuit depth by directly measuring the corresponding Pauli operator on the quantum hardware to get the finite sample estimators $\widehat{P_k}$ (here and in the following we will use a wide hat to indicate sample estimators) and combine them to form
\begin{equation}
\label{eq:lcu}
\widehat{O_A} = \alpha_0  +  \sum_{k=1}^L \alpha_k \widehat{U_k} = \alpha_0  +  \sum_{k=1}^L \alpha_k e^{i\theta_k} \widehat{P_k}\;,
\end{equation}
which converges to $\langle O\rangle$ in the infinite measurement limit. In particular, if we perform $M$ measurement for every one of the $L$ terms in Eq.~\eqref{eq:lcu} the variance of this estimator is given simply by
\begin{equation}
\label{eq:mu_var}
Var[\widehat{O_A}] = \sum_{k=1}^{L} \alpha_k^2 Var[\widehat{P_k}] = \sum_{k=1}^{L} \alpha_k^2 \frac{1-\widehat{P_k}^2}{M} \;,
\end{equation}
due to the independence of the separate measurements. Note that this implies that in general $Var[\widehat{O_A}]\neq Var[O]$ and in fact Eq.~\eqref{eq:mu_var} may be large even for eigenstates of $O$.
Using Eq.~\eqref{eq:mu_var} we can estimate the total number of measurement $N_{tot}=LM$ required to evaluate $\widehat{O_A}$ with precision $\epsilon$ to be
\begin{equation}
\label{eq:Ntot_bound}
N_{tot} = \frac{L}{\epsilon^2} \sum_{k=1}^{L} \alpha_k^2\left[1-\widehat{P_k}^2\right] \leq \frac{L\|\overline{O_T}\|_2^2}{\epsilon^2}\;,
\end{equation}
where we defined
\begin{equation}
\label{eq:qnorm}
 \|\overline{O_T}\|_q \equiv \left(\sum_{k=1}^L \alpha_k^q\right)^{1/q}\quad \text{for}\quad q\geq1\;.
\end{equation}
Different strategies have been proposed in the literature to reduce the scaling of operator averaging shown in Eq.~\eqref{eq:Ntot_bound}. For instance, in situations like quantum chemistry where a large number of coefficients $\alpha_k$ have possibly a small magnitude, one can use efficient truncation schemes~\cite{McClean2014,McClean2016} to improve the performance considerably. A complementary approach is to group the $L$ terms into $G$ groups of operators which can be measured together in a single experiment~\cite{McClean2016,Kandala2017,Izmaylov2019}, if the newly introduced correlations are not too large this approach can allow again a great reduction in the number of measurements. In general however we still expect $G$ to scale with system size (for quantum chemistry applications see eg.~\cite{McClean2014}). 
We could also choose the number of measurements to be performed for $k$-th term to depend on the magnitude of the expansion coefficients $\alpha_k$ as proposed in~\cite{Romero2018}:
\begin{equation}
M_k \propto \frac{\alpha_k}{\|\overline{O_T}\|_1}\;,
\end{equation}
this leads in turn to the estimate
\begin{equation}
\label{eq:Ntotp_bound}
N'_{tot} = \frac{\|\overline{O_T}\|_1}{\epsilon^2} \sum_{k=1}^{L} \alpha_k \left[1-\widehat{P_k}^2\right]
\end{equation}
and in this way obtaining a better bound for $N_{tot}$ by a factor $\|\overline{O_T}\|_1^2/(L \|\overline{O_T}\|_2^2) \leq 1$. For our nuclear physics problem this strategy produced actually a small increase in the cost since the magnitude of the variance of $k$-th term does not necessarily correlate with the magnitude of coefficient $\alpha_k$. 
By also using the asymptotic improvement from this adaptive variant, we will consider
\begin{equation}
\label{eq:Ns_eps}
N_A(\epsilon) = \frac{\|\overline{O_T}\|_1^2}{\epsilon^2} \lesssim \frac{L \|\overline{O_T}\|_2^2}{\epsilon^2}
\end{equation}
as an estimate for the number of repetition needed to obtain a precision $\epsilon$ with operator averaging. 

As expected we find that the estimator of Eq.~\eqref{eq:lcu} shows in all cases the usual shot noise behavior for small errors $N_{tot}=\mathcal{O}(1/\epsilon^2)$ but its explicit dependence on the operator norm of $O_T$ (which is  a lower bound for $\|\overline{O_T}\|_1$) can be unfavorable when the expectation value $\langle O\rangle$ becomes too small. Introducing the ratio 
\begin{equation}
\label{eq:ratio}
R_O = \frac{\lvert \langle O\rangle\rvert}{\|\overline{O_T}\|_1}\leq\frac{\|\overline{O}\|_1}{\|\overline{O_T}\|_1} \equiv R^{max}_O
\end{equation}
we can express the number of shots $N_{tot}$ in terms on the relative error $\epsilon_r=\epsilon/\lvert\langle O\rangle\rvert$ as
\begin{equation}
\label{eq:Ns_erel}
N_A(\epsilon_r) = \frac{1}{\epsilon_r^2 R_O^2}\geq\frac{1}{\epsilon_r^2}\left(\frac{\lambda_{max}-\alpha_0}{\lvert\langle O\rangle\rvert}\right)^2
\end{equation}
which makes explicit the quadratic dependence of the classical effort (the number of repetitions) with the ratio between the largest eigenvalue $\lambda_{max}$ of the target operator $O$ and its expectation value in the state $\ket{\Psi}$. 

The scheme we propose in the next section can be advantageous whenever the ratio $R_O$ becomes excessively small by providing a scheme with $N_{tot}$ independent of $R_O$ in the special case of eigenvalue estimation and possibly well performing in general (see condition Eq.~\eqref{eq:cond}). This could be important in a large system when $\|O_T\|$ is extensive (indeed for applications in quantum simulation we expect $\|O_T\|=\mathcal{O}\left(poly(n)\right)$ in the number of qubits, see e.g.~\cite{McClean2014}) or simply because the expectation value we are after is much smaller than the largest eigenvalue of $O_T$ like for the ground state of the deuteron with hard-core potentials~\cite{Wiringa1995,Carlson2015} studied in Sec.~\ref{sec:deuteron}.

\section{Expectation values from single step phase estimation}
\label{sec:newmethod}

The widespread use of the direct algorithm described in the previous section comes from the appealing property of minimizing the required quantum resources by not requiring any additional quantum operation on the qubits. This is especially important for NISQ era devices where coherence time and noise will limit the attainable circuit depth~\cite{Preskill2018}.
As mentioned in the introduction, algorithms based on full fledged quantum phase estimation, like the one described in~\cite{Knill07}, will possibly allow to approach Heisenberg limited scaling $\epsilon=\mathcal{O}(1/N)$ as a function of the number of measurements $N$. The price for this is a circuit depth that scales as $C_D=\mathcal{O}(1/\epsilon)$ making its implementation challenging on noisy devices with circuit depth $C\lesssim100$~\cite{Preskill2018}. In this section we show how to effectively use a single step of time evolution to obtain a large efficiency gain in measurement number with respect to the operator averaging method, while enjoying short circuit depths $C_{sQPE}=\mathcal{O}(\epsilon^\alpha)$ with $\alpha\geq0$.

Similarly to the technique of~\cite{Knill07}, we can use the small time expansion of the imaginary part of the expectation value of the time evolution unitary $U_\tau=e^{i\tau O}$ on the state $\ket{\Psi}$ (cf. Eq.~\eqref{eq:smallt_exp})
\begin{equation}
\label{eq:sin_exp}
\langle sin\left(\tau O\right)\rangle = \tau \langle O\rangle - \frac{\tau^3}{6}\langle O^3\rangle + \mathcal{O}\left(\tau^5\|\overline{O}\|_1\right)
\end{equation}
to extract the expectation value $\langle O\rangle$. In particular for any Hermitian observable $O$ we can consider the following standard circuit (cf.~\cite{Cleve98,Ortiz01}):
\begin{equation}
\label{eq:circuitcH}
\Qcircuit @C=1em @R=.7em {
\ket{0}\quad\quad& \gate{H} &\ctrl{1} & \gate{S}&\gate{H}&\qw\\%&\meter \\
\ket{\Psi}\quad\quad&\qw& \gate{e^{i\tau O}} & \qw& \qw& \qw\\%&\qw\\
}
\end{equation}
The imaginary part of $Tr\left[e^{i\tau O}\rho\right]$ can be extracted by measuring the ancilla along the z axis
\begin{equation}
\label{eq:zanc}
\begin{split}
\langle Z_{a} \rangle(\tau) &= p_0(\tau)-p_1(\tau)\\
&=- \bra{\Psi}sin\left(\tau O\right)\ket{\Psi}
\end{split}
\end{equation}
where $p_0$ ($p_1$) are the probabilities of measuring the ancilla in the $\ket{0}$ ($\ket{1}$) state. 
In practice we need to estimate the expectation value Eq.~\eqref{eq:zanc} by performing $N$ independent measurements and computing their average $\widehat{Z_a}(\tau)$. For ease of use, from here on we will call the process of extracting $\langle O\rangle$ from a polynomial fit of a set of estimates of Eq.~\eqref{eq:zanc} for different times $\tau$ as sQPE for single step quantum phase estimation.

It is important to note now that, due to the presence of a bias for finite values of $\tau$ (coming from the necessity to truncate the series expansion in Eq.~\eqref{eq:sin_exp}),  sQPE has a worse asymptotic scaling $N=\mathcal{O}\left(\epsilon^{-(2+\kappa)}\right)$ with $\kappa>0$. Despite this efficiency loss in the asymptotic $\epsilon\to 0$ limit, we will show in the next subsections that the constant factors can be very small when the ratio $R_O\ll1$. provided certain conditions are met (see Eq.~\eqref{eq:cond} and Sec.~\ref{sec:pcond}). For instance in the situation when we prepare an eigenstate of $O$ and we want to estimate it's eigenvalue $\lambda_\Psi$, a bound $R_O\leq 1\%$ is sufficient to guarantee an advantage for the simplest possible version of sQPE (see Sec.~\ref{sec:lin}) up to a respectably small relative error $\epsilon_r=\mathcal{O}(10^{-5})$.

We next describe the general high-order sQPE in an idealized setting and in the following two subsection we propose practical implementation for the two lowest order sQPE algorithms where the truncation of Eq.~\eqref{eq:sin_exp} occurs at either the linear (Sec.~\ref{sec:lin}) or the cubic term (Sec.~\ref{sec:cub}). For now we will focus our attention on the classical resources (the number of measurements that needs to be performed) and postpone the discussion of the cost associated with the implementation of the (controlled) time evolution unitary $U_\tau=e^{i\tau O}$ in Eq.~\eqref{eq:circuitcH} to section Sec.~\ref{sec:time_evo}. We can already anticipate that the gate count will be low since the total evolution time $\tau$ has to be kept small enough to minimize the effect of the bias.

\subsection{General scaling}
\label{sec:general_discussion}
Let's start by considering the idealized case where we know the coefficients $m_k=\langle O^{2k+1}\rangle$ for $k={1,\dots,K}$ (in practice we will need to estimate these resulting in a sub-optimal algorithm). We can now use the Taylor expansion in $\tau$ of Eq.~\eqref{eq:zanc} to construct the following biased estimator for the expectation value:
\begin{equation}
\label{EKtau}
O_K(\tau) = -\frac{1}{\tau}\left(\langle Z \rangle_{a}(\tau)+\sum_{k=1}^{K-1} \tau^{2k+1}\frac{(-1)^km_k}{(2k+1)!}\right)\;.
\end{equation}
Here the bias comes from neglecting higher order terms with $k\geq K$ in the expansion
\begin{equation}
\begin{split}
B_K(\tau) &= O_K(\tau) - \langle O \rangle= -\sum_{k=K}^\infty \tau^{2k}\frac{(-1)^km_k}{(2k+1)!}\; ,
\end{split}
\end{equation}
and its magnitude can bound by using Lagrange’s Remainder theorem:
\begin{equation}
\label{eq:ubound}
\lvert B_K(\tau)\rvert\leq \tau^{2K} \frac{\lvert m_K\rvert}{(2K+1)!}\;.% \equiv B^u_K(\tau)\; .
\end{equation}

Due to the presence of this bias for any finite value of $\tau$ we choose to characterize the deviations of our estimator to the exact expectation value by means of the Mean Squared Error (MSE) defined as:
\begin{equation}
\label{eq:mse}
\epsilon_M^2(\tau,K) = Var\left[\widehat{O_K}(\tau)\right] + B_K(\tau)^2\;,
\end{equation}
where, as before, we denote with $\widehat{O_K}(\tau)$ a finite population estimator of Eq.~\eqref{EKtau}. The expected total number of measurements required to achieve a final precision (meaning MSE) target $\epsilon$ can then be estimated as was done before in Sec.~\ref{sec:SM}. In particular for any $\epsilon >B_K(\tau)$ we have the following
\begin{equation}
N_{tot}=\frac{1}{\tau^2}\frac{1-\widehat{Z_a}(\tau)^2}{\epsilon^2-B_K(\tau)^2}\;,
\end{equation}
where we used $Var[\widehat{O_K}(\tau)]=(1-\widehat{Z_a}(\tau)^2)/(N\tau^2)$ with $N$ the size of the population used to estimate $\widehat{Z_a}(\tau)$. This estimate is minimized with the choice $\tau=\tau_{\rm opt}$ with
\begin{equation}
\label{eq:time_step}
\tau_{\rm opt}=\left(\frac{(2K+1)!}{\sqrt{2K+1}}\frac{\epsilon}{\lvert m_K\rvert}\right)^{\frac{1}{2K}}%\nonumber\\
%&=&\left(\frac{(2K+1)!}{\sqrt{2K+1}}\right)^{\frac{1}{2K}}\left(\frac{\epsilon}{\rvert m_K\rvert}\right)^{\frac{1}{2K}}\;.
\end{equation}
which leads to the following estimate for the total number of measurements needed for sQPE at order $K$:
\begin{equation}
\label{eq:NT}
N_{sQPE}(K,\epsilon) = \frac{|m_K|^{1/K}}{\epsilon^{2+1/K}} f(K)
\end{equation}
where we have defined
\begin{equation}
f(K) = \frac{2K+1}{2K}\left( \frac{\sqrt{2K+1}}{(2K+1)!}\right)^{\frac{1}{K}}\;,
\end{equation}
and as promised the shot noise regime $N_{sQPE}=\mathcal{O}(1/\epsilon^2)$  is recovered only asymptotically for large values of $K$. As anticipated then, for a sufficiently small error we expect the Operator Averaging method of Sec.~\ref{sec:SM} to outperform the scheme presented here. The advantage of sQPE is in the possibility of having a much weaker dependence of the expectation value to norm ratio $R_O$. To see this let us first rewrite Eq.~\eqref{eq:NT} using the relative error instead
\begin{equation}
\label{eq:NT_erel}
N_{sQPE}(K,\epsilon_r) = \frac{f(K)}{\epsilon_r^{2+1/K}} \frac{\lvert\langle O^{2K+1}\rangle\rvert^{1/K}}{\lvert\langle O\rangle\rvert^{2+1/K}} \;,
\end{equation}
where we simply used the definition of $m_k$. Consider now the special case of eigenvalue estimation where $\ket{\Psi}$ is an eigenvector of $O$ with eigenvalue $\lambda_\Psi$ that we want to compute. In this limit the ratio of expectation values on the right-hand side of Eq.~\eqref{eq:NT_erel} is just $1$ and the total number of measurement required by sQPE is completely independent on system considered, and in particular it does not depend on $R_O$. In the more general case where $\ket{\Psi}$ is not a single eigenvector we will need an additional ingredient to assess the performance of sQPE: a tight upper bound on the bias Eq.~\eqref{eq:ubound} or equivalently of the moment $m_K$. For this purpose let us first introduce $\Gamma_K$ as an upperbound of the following ratio
\begin{equation}
\label{eq:gamma_ratio}
\frac{\lvert m_K\rvert}{\|\overline{O}\|^{2K+1}_1} = \frac{\lvert \langle O^{2K+1}\rangle\rvert}{\|\overline{O}\|^{2K+1}_1} \leq \Gamma_K \leq 1\;,
\end{equation}
where we remind that $\|\overline{O}\|_1 = \lvert\alpha_0\rvert + \|\overline{O_T}\|_1$. We can now use $\Gamma_K$ to bound the number of measurements as
\begin{equation}
\label{eq:NT_erel_bound}
N_{sQPE}(K,\epsilon_r) \leq \frac{f(K)}{\epsilon_r^{2+1/K}} \frac{\Gamma_K^{1/K}}{R^{2+1/K}_O} \left(\frac{\|\overline{O}\|_1}{\|\overline{O_T}\|_1}\right)^{2+1/K}\;,
\end{equation}
and this could be smaller than Eq.~\eqref{eq:Ns_erel} when $\Gamma_K\ll1$. To be more quantitative, the sQPE estimator Eq.~\eqref{EKtau} will become efficient when $N_{sQPE}(K,\epsilon_r)\leq N_A(\epsilon_r)$ at the desired relative accuracy $\epsilon_r$. Using Eq.~\eqref{eq:Ns_erel} and Eq.~\eqref{eq:NT_erel}, this condition can be written equivalently as
\begin{equation}
\label{eq:cond}
R_O\geq \frac{f(K)^K}{\epsilon_r} \frac{\lvert\langle O^{2K+1}\rangle\rvert}{\|\overline{O_T}\|_1^{2K+1}}\;,
\end{equation}
which can be turned in the following sufficient condition
\begin{equation}
\label{eq:cond_suff}
R_O\geq \frac{f(K)^K}{\epsilon_r} \Gamma_K \left(1+\frac{\lvert\alpha_0\rvert}{\|\overline{O_T}\|_1}\right)^{2K+1}\;,
\end{equation}
where we reintroduced the upperbound $\Gamma_K$ described above and wrote explicitly the dependence on the coefficient $\alpha_0$ in Eq.~\eqref{eq:op_exp}. We will discuss in detail how to check if the condition Eq.~\eqref{eq:cond} is satisfied in practical application in Sec.~\ref{sec:pcond} while for now we focus on the simpler situation of eigenvalue estimation mentioned above. In this case the right hand side of Eq.~\eqref{eq:cond} takes a simple form and the full inequality can be written as:
\begin{equation}
\label{eq:cond_eig}
\frac{\lvert\lambda_\Psi\rvert}{\|\overline{O_T}\|_1} \leq \frac{\epsilon_r^{1/2K}}{\sqrt{f(K)}}\;.
\end{equation}
We now need only a reasonably tight upperbound $\lambda_u$ of $\lvert\lambda_\Psi\rvert$ to judge when the condition Eq.~\eqref{eq:cond_eig} is satisfied. This requirement is rather loose in practice since the left-hand side is bounded from above by $\lambda_{max}/(\lambda_{max}-\alpha_0)\approx \mathcal{O}(1)$ while the right hand side becomes large quickly as a function of $K$.

\begin{figure}
    \centering
    \includegraphics[scale=0.3]{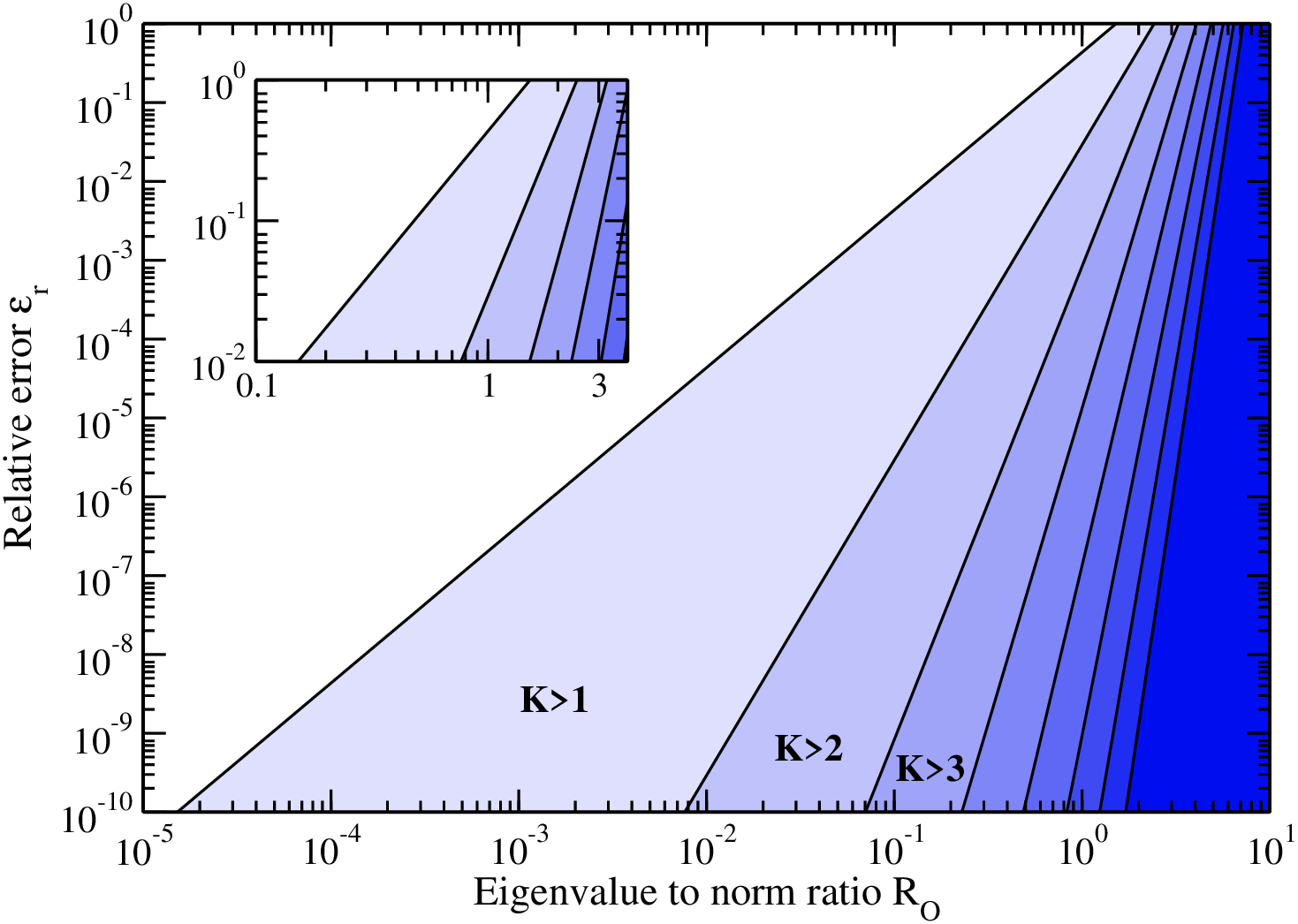}
    \caption{The solid black curves indicate the set of points where Eq.~\eqref{eq:cond_eig} is satisfied with the equality for different choices of the sQPE order $K$ with an ordering from top left to bottom right: the first solid line refers to $K=1$ while the second is for $K=2$ and the shaded region in between indicates the region where the linear method is no longer sufficient and we need to employ sQPE with $K>1$. The innermost (darkest) region is bordered by the $K=8$ line.
    With the inset we are zooming in the top-right corner of the main plot.}
    \label{fig:econdB}
\end{figure}

In order to visualize this effect, we plot in Fig.~\ref{fig:econdB} the minimum relative error $\epsilon_r$ that can be achieved for a fixed value of the eigenvalue ratio $R_O$ using the condition from Eq.~\eqref{eq:cond_eig}. As this will depend on the chosen order $K$, in the figure we report the boundaries starting from $K=1$ in the top left corner up to $K=8$ in the bottom right. In particular this means that to achieve lower error rates than the first solid curve we will need to use sQPE with $K>1$ in order to ensure the condition Eq.~\eqref{eq:cond_eig} is satisfied (ie. sQPE could provide an advantage over the Operator Averaging method). Conversely the darkest and innermost region in Fig.~\ref{fig:econdB} is accessible only for $K>8$.

We can deduce a number of interesting conclusions from this figure. For instance, we can see that if we can place an upperbound of $R_O$ smaller than $\approx10^{-2}$, then sQPE with $K=2$ will be more efficient than Operator Averaging down to extremely small relative errors $\epsilon_r\approx10^{-9}$. This is a major improvement from the limit $\epsilon_r\approx10^{-4}$ achievable with the linear method of Sec.~\ref{sec:lin}.
Furthermore, as we can clearly see in the inset, for target relative error at the $1\%$ level, the linear method can be used effectively up to $R_O\approx0.1$ while by increasing the order to cubic (i.e. $K=2$) we can push this up to $R_O\approx0.8$.
Given these observations and the increasing difficulty in implementing sQPE efficiently for large $K$, it's likely that for many practical situations sQPE with $K=1,2$ will be sufficient to achieve a substantial speedup in terms of the number of measurements required to estimate the eigenvalue $\lambda_\Psi$.
We will provide a similar scaling analysis for the more general problem of expectation value estimation in Sec.~\ref{sec:pcond} (see Fig.~\ref{fig:ubound}).

It is now time to come back to the problem of performing the polynomial fit needed in Eq.~\eqref{EKtau} in the realistic case where we do not know the high-order coefficients $m_k$. For the last coefficient with $k=K$ we can easily use a reasonably tight upperbound $\Gamma_K$ to manage the influence of the bias in  Eq.~\eqref{eq:ubound} as we did before. This allows for a complete algorithm in the simplest case $K=1$ achieving the scaling of Eq.~\eqref{eq:NT_erel_bound}. 
In higher order algorithms we need to estimate the higher order contributions for $k<K$ by collecting data at different values of the time-step $\tau$ and performing a non-linear fit. If no other information is available, we will need at least $K$ values of $\tau$ to properly perform the reconstruction (eg. using the expansion proposed in~\cite{Knill07}).

In the following sections we describe an implementation of the simple $K=1$ (linear) algorithm and a more efficient scheme that adaptively finds the optimal pair $(\tau_a,\tau_b)$ of time parameters for the $K=2$ (cubic) case.

\subsection{Linear Algorithm}
\label{sec:lin}

The simplest case is where we neglect the cubic terms in the expansion of the $sin$ so that our estimator Eq.~\eqref{EKtau} becomes
\begin{equation}
\label{eq:eest}
O_1(\tau)= \frac{1}{\tau}\bra{\Psi}sin\left(\tau H\right)\ket{\Psi} = -\frac{1}{\tau}\langle Z \rangle_a(\delta) \;.
\end{equation}
In the linear case the optimal time-step Eq.~\eqref{eq:time_step} is
\begin{equation}
\label{eq:tau_lin}
\tau_{opt} = \sqrt{\frac{6}{\sqrt{3}}\frac{\epsilon}{\lvert\langle O^3\rangle\rvert}}=\sqrt{\frac{6}{\sqrt{3}}\frac{\epsilon_r\lvert\langle O\rangle\rvert}{\lvert\langle O^3\rangle\rvert}}\;,
\end{equation}
and now, using the estimate of Eq.~\eqref{eq:NT_erel}, we can estimate the number of measurements needed as
\begin{equation}
N_{sQPE}(1,\epsilon_r) =\frac{1}{\epsilon_r^3} \frac{\sqrt{3}}{4}\left\lvert\frac{\langle O^3 \rangle }{\langle O \rangle^3}\right\rvert\longrightarrow \frac{1}{\epsilon_r^3} \frac{\sqrt{3}}{4}\;,
\end{equation}
where the limit on the right hand side holds when we prepare $\ket{\Psi}$ in any eigenvector of $O$. Note that in this limit, the resources required for a given target relative accuracy $\epsilon_r$ are completely independent on the chosen operator $O$ or even eigenvalue $\ket{\Psi}$. For instance at the $1\%$ level we have
\begin{equation}
\label{eq:lin_1perc}
N_{sQPE}(1,\epsilon_r=0.01) = \frac{\sqrt{3}}{4}\times10^6\approx 4.3 \times 10^5\;.
\end{equation}
However, in order to get an advantage with sQPE for this situation, the inequality of Eq.~\eqref{eq:cond_eig} has to be satisfied, and that one does depend on the system details.

The problem now is that in general situations we will not be able to calculate $\tau_{opt}$ without at least an approximate estimate for the wanted expectation value $\langle O\rangle$, a good bound $\Gamma_K$ on the bias is in fact not sufficient. Even in the simpler case of eigenvalue estimation we can only use an upperbound $\lambda_u$ for the absolute value of eigenvalue to compute the approximation
\begin{equation}
\label{eq:atau_lin}
\widetilde{\tau}_{opt}=\sqrt{\frac{6}{\sqrt{3}}\frac{\epsilon_r}{\lambda^2_u}}\leq\tau_{opt}\;,
\end{equation}
and this will cause the total number of measurement required to increase:
\begin{equation}
\label{eq:lin_app}
\widetilde{N}_T(1,\epsilon_r) = \left(\frac{\lambda_u}{\lvert\lambda_\Psi\rvert}\right)^{3/2}N_{sQPE}(1,\epsilon_r)\;.
\end{equation}
This is somewhat acceptable in the eigenvalue estimation case as an upperbound for the eigenvalue is used initially to check if sQPE is at all convenient using condition Eq.~\eqref{eq:cond_eig}. The relatively strong dependence of $\widetilde{N}_T(1,\epsilon_r)$ on the value of $\lambda_u$ and the difficulty to obtain the optimal time-step in the general case are two of the main problems of the linear algorithm just described and strong motivations for developing the self-consistent algorithm described in the next section.

\begin{figure}
    \centering
    \includegraphics[scale=0.3]{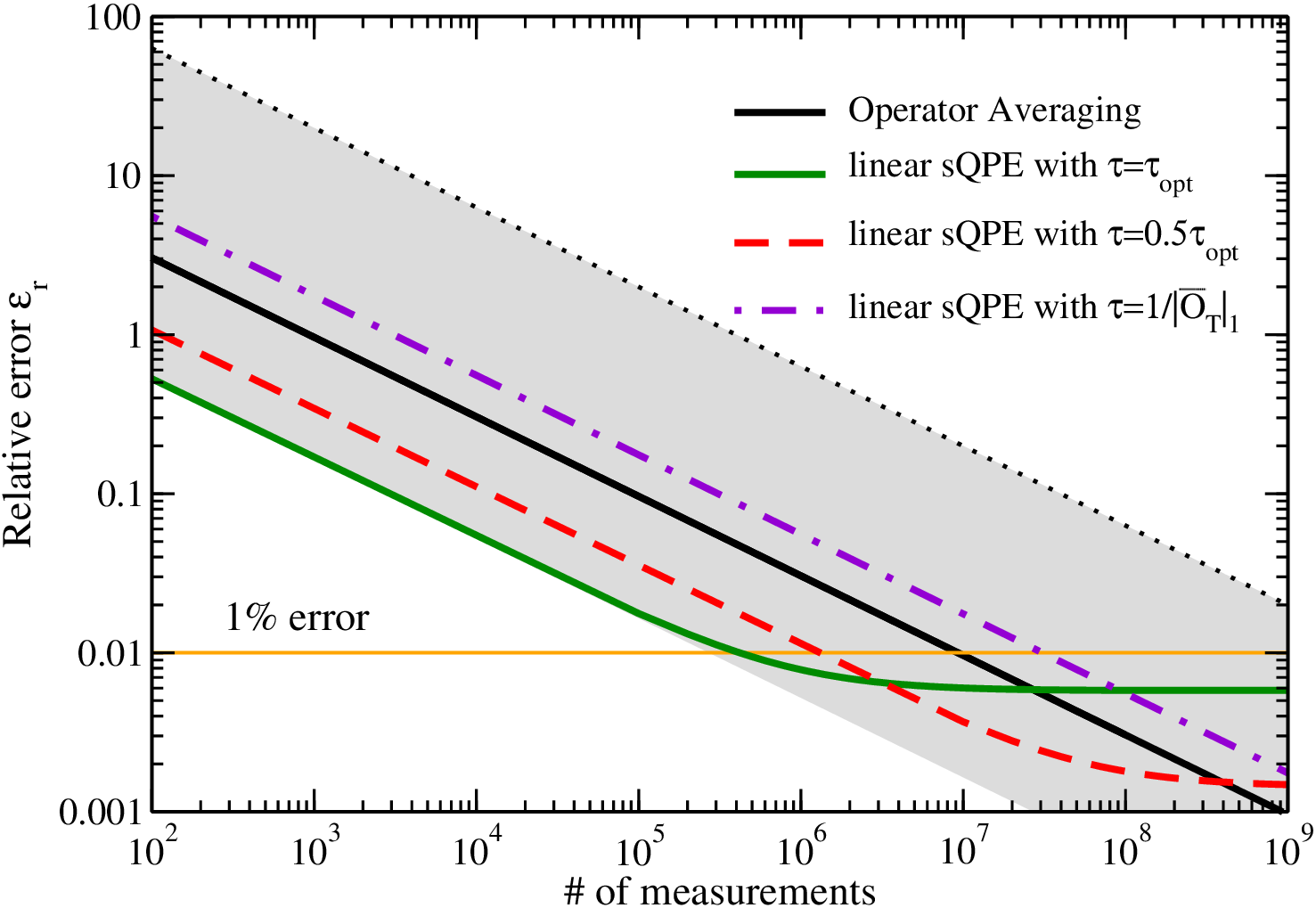}
    \caption{Results of numerical simulations of the linear algorithm explained in the text together with the operator averaging method of Sec.~\ref{sec:SM}. The grey band indicates the expected variation caused by different choices for the time step $\tau$ in the linear algorithm. The orange line indicates an indicative $\epsilon_r=1\%$ target error threshold.}
    \label{figB}
\end{figure}

Before moving to the cubic algorithm, we want to further illustrate the sensitivity of the linear method on the particular choice for the time step $\tau$ by anticipating some results from the deuteron model described in detail in Sec.~\ref{sec:deuteron}. In Fig.~\ref{figB} we show with a black solid line the analytical estimate (cf. Eq.~\eqref{eq:Ntot_bound}) for the error scaling of the operator averaging method of Sec.~\ref{sec:SM}
\begin{equation}
\label{eq:err_oa}
\epsilon = \sqrt{\frac{L}{N_{tot}} \sum_{k=1}^{L} \alpha_k^2\left[1-\widehat{P_k}^2\right] }\;,
\end{equation}
as a function of the total number of measurements $N_{tot}$. As mentioned in Sec.~\ref{sec:SM} above the simple adaptive scheme of Eq.~\eqref{eq:Ntotp_bound} produces a slightly worse performance ($\approx10\%$ larger constant factor, not shown) than the naive operator averaging method.

For sQPE with $K=1$ (or linear method) we show instead the mean squared error
\begin{equation}
\label{eq:mse_lin}
\epsilon_M(\tau,1) = \sqrt{\frac{1-\langle Z_a\rangle(\tau)^2}{\tau^2N}+\tau^4 \frac{\lvert\langle O^3\rangle\rvert^2}{36}}\;,
\end{equation}
achievable with different time-steps. The green line corresponds to sQPE with $K=1$ using the optimal choice for the time step Eq.~\eqref{eq:tau_lin} while the red dashed curve shows the detrimental effect of using a worse upperbound $\lambda_u=2\lvert\lambda_\Psi\rvert$ resulting in $\tau=\tau_{opt}/2$. In both cases sQPE provides an important speedup over operator averaging but this advantage is very fragile. The purple dot-dashed curve shows results obtained using the smallest time-step which could reasonably provide an advantage: the time step for which the variance in Eq.~\eqref{eq:mse} equals an upperbound of the variance of the operator averaging estimator: $1/\tau=\|\overline{O_T}\|_1$. The performance of operator averaging can be no worse than that, and indeed we see in Fig.~\ref{figB} that linear sQPE with this time-step requires $\approx3$ times more measurements than the original scheme. The grey band spans the whole region covered by linear order algorithms with varying time-steps, in particular the upperbound (shown as dotted black line in Fig.~\ref{figB}) corresponds to the worse possible choice for the eigenvalue upperbound $\lambda_u=\lambda_{max}$.

The linear algorithm is simple to implement and can provide already important efficiency gains over Operator Averaging whenever we have the ability to make a good choice for the time-step parameter $\tau$. We will now show how we propose to tackle this issue by using the next order sQPE algorithm corresponding to $K=2$.

\subsection{Cubic algorithm}
\label{sec:cub}
In order to use the estimator Eq.~\eqref{EKtau} for $K=2$ we need to be able to estimate $m_1=\lvert\langle O^3\rangle\rvert$. In our implementation we achieve this by computing $\langle Z\rangle_a(\tau)$ for two different values of the time step and use these to extract both $\langle O\rangle$ and $m_1$ using a cubic fit.
Given a pair of time steps $(\tau_a,\tau_b)$ the outcome of $M$ independent measurements over of the projector $\Pi_a=\rvert0\rangle\langle0\lvert$ is described by a pair of binomial random variables $X_a\sim B(M,P_a)$ and $X_b\sim B(M,P_b)$ with probabilities given by
\begin{equation}
\label{eq:pair_prob}
\begin{split}
&P_{a/b} = \frac{1-\langle\Psi\lvert sin(\tau_{a/b} O)\rvert\Psi\rangle}{2} \\
&=\frac{1-\tau_{a/b} \langle\Psi\lvert O\rvert\Psi\rangle+\frac{\tau^3_{a/b}}{6}\langle\Psi\lvert O^3\rvert\Psi\rangle}{2}+\mathcal{O}\left(\tau_{a/b}^5\right)\;.
\end{split}
\end{equation}
For small values of the time steps we can approximate these distributions with
\begin{equation}
\widetilde{P}_{a/b}(\mu,\eta) = \frac{1-\tau_{a/b}\; \mu+\frac{\tau_{a/b}}{6}\eta}{2} \;.
\end{equation}
Estimators for the two parameters $\mu$ and $\eta$ can be obtained by extremizing the likelihood $ L(X_a,X_b | \mu,\eta,\tau_a,\tau_b)$ to observe a particular realization $(X_A,X_B)$ given the distribution parameters $(\mu,\eta,\tau_a,\tau_b)$:
\begin{multline}
\label{eq:likelihood}
 L(X_a,X_b | \mu,\eta,\tau_a,\tau_b) \propto \widetilde{P}^{X_a}_a(\mu,\eta) \left(1-\widetilde{P}_a(\mu,\eta)\right)^{M-X_a}\\
\times\widetilde{P}^{X_b}_b(\mu,\eta) \left(1-\widetilde{P}_b(\mu,\eta)\right)^{M-X_b}\;.
\end{multline}
The resulting maximum likelihood estimators are:
\begin{equation}
\label{eq:mle_estim}
\mu_{mle} = c^\mu_{ab}\left[\frac{\tau_a^2}{\tau_b}\left(1-2\frac{X_b}{M}\right)-\frac{\tau_b^2}{\tau_a}\left(1-2\frac{X_a}{M}\right)\right]
\end{equation}
\begin{equation}
\eta_{mle} = c^\eta_{ab}\left[\tau_a\left(1-2\frac{X_b}{M}\right)-\tau_b\left(1-2\frac{X_a}{M}\right)\right]
\end{equation}
where the time-step dependent coefficients are
\begin{equation}
c^\mu_{ab} = \frac{1}{\tau_a^2-\tau_b^2}\quad\text{and}\quad c^\eta_{ab} = \frac{6}{\tau_a\tau_b(\tau_a^2-\tau_b^2)}\;.
\end{equation}

We can estimate the variance of these estimators by computing the inverse of the Fisher information matrix
\begin{equation}
\label{eq:fisher}
I(\mu,\eta)_{i,j}= - \mathbb{E}\left[\frac{\partial^2 log\left(L(X_a,X_b\vert\mu,\eta,\tau_a,\tau_b)\right) }{\partial i\partial j}\right]\;,
\end{equation}
where the derivatives are taken over $\{i,j\}=\{\mu,\eta\}$ and the parametric dependence of $I(\mu,\eta)$ on the two time-steps $\tau_{a/b}$ has been suppressed for clarity. The results are
\begin{equation}
Var\left[\mu_{mle}\right] = \frac{4}{M}\frac{\tau_a^6P_b\left(1-P_b\right)+\tau_b^6P_a\left(1-P_a\right) }{\tau_a^2\tau_b^2\left(\tau_a^2-\tau_b^2\right)^2}
\end{equation}
\begin{equation}
Var\left[\eta_{mle}\right] = \frac{144}{M}\frac{\tau_a^2P_b\left(1-P_b\right)+\tau_b^2P_a\left(1-P_a\right) }{\tau_a^2\tau_b^2\left(\tau_a^2-\tau_b^2\right)^2}
\end{equation}
where the probabilities $P_{a/b}$ will need to be estimated using only a finite sample. In this work we used the Bayesian estimators
\begin{equation}
\widehat{P_a} = \frac{X_a+1}{M+2}\quad\quad\quad \widehat{P_b} = \frac{X_b+1}{M+2}\;,
\end{equation}
obtained using a slightly informative Beta prior with $\alpha=\beta=1$, but in general any accurate sample estimator $\widehat{P_{a/b}}$ of the true probabilities $P_{a/b}$ will do.

Note that estimators of the variance obtained in this way are in principle accurate only in the limit of large statistics $M\gg1$, but in practice we found their use to be reasonable for the application studied in this work. Since the adaptive algorithm we describe below relies on their quality, further work on the construction of better (ie. more robust to noise or more rapidly converging) estimators of the fluctuations of $\mu$ and $\eta$ may prove useful.

In addition to these statistical sources of error we also have a bias coming from the approximation $P_{a/b}\to\widetilde{P}_{a/b}$. This can be estimated to be
\begin{equation}
\label{eq:bias_cub_e}
\begin{split}
&B_\mu(\tau_a,\tau_b)=  \mathbb{E}\left[\mu_{mle} - \mu\right]\\
&=c^\mu_{ab}\left[\frac{\tau_a^2}{\tau_b} \left(1-2P_b\right)-\frac{\tau_b^2}{\tau_a} \left(1-2P_a\right)\right]-\mu\\
&=c^\mu_{ab}\left[\frac{\tau_a^2}{\tau_b} \langle sin(\tau_b O)\rangle-\frac{\tau_b^2}{\tau_a}\langle sin(\tau_a O)\rangle\right]-\mu\;.
\end{split}
\end{equation}
A useful upperbound can be obtained by noticing that $B_\mu(\tau_a,\tau_b)$ is obtained from the remainder of the Taylor expansion in Eq.~\eqref{eq:pair_prob} as
\begin{equation}
B_\mu(\tau_a,\tau_b) = \langle\Psi\lvert R_5 \rvert\Psi\rangle 
\end{equation}
where the remainder operator $R_5$ is defined as
\begin{multline}
\label{eq:r5}
R_5 = \frac{c^\mu_{ab}O^5}{24}\int_0^1 dt (1-t)^4\bigg[\tau_a^2\tau_b^4 cos(t\tau_bH)\\
-\tau_b^2\tau_a^4cos(t\tau_aH)\bigg]\;,
\end{multline}
as follows directly from the integral representation of the remainder of a Taylor series. We can now bound the bias in the cubic algorithm using for instance
\begin{equation}
\label{eq:bias_cub}
 \lvert B_\mu(\tau_a,\tau_b) \rvert \leq \frac{ \lvert \langle\Psi\lvert O^5 \rvert\Psi\rangle\rvert}{120}\tau_a^2\tau_b^2\frac{\tau_a^2+\tau_b^2}{\lvert \tau_a^2-\tau_b^2 \rvert}\;.
\end{equation}
Achieving a tight bound for the bias is generally important as it controls the final efficiency of the method. As for the linear method above, for now we will focus on the special case of eigenvalue estimation while leaving the discussion on how to obtain practical upperbounds in more general situations in Sec.~\ref{sec:pcond}. 
In the calculations performed in our work we found a weak dependence of the computational effort with the particular choice of estimator for the bias (see Fig.~\ref{fig:biasB}) and we will discuss the different options we used in Sec.~\ref{sec:deuteron}.

In the next subsection we present our strategy to determine the time steps for the cubic algorithm.

\subsubsection{Optimal determination of the times steps}
As was pointed out at the end of Sec.~\ref{sec:lin} one of the major drawbacks of the linear algorithm is its sensitivity to the choice of the time-step $\tau$. For the cubic algorithm with solve this issue by using ideas form Optimal Design~\cite{Chernoff_ODbook,Loredo2004}.
Optimal Design (OD) techniques have been used in a variety of applications to quantum computing ranging from quantum tomography~\cite{Kosut2004,Nunn2010,Huszar2012}, to parameter estimation~\cite{Ballo2011,Granade2012}, to quantum-gate synthesis~\cite{Schulte2005}.
The general underlying idea in OD for parameter estimation is to try to optimize some, possibly unconstrained, hyper-parameters of an experiment (eg. the pair of time steps ($\tau_a,\tau_b)$ to be used in the cubic sQPE) in order to minimize an estimator for the error in the parameter we want to estimate.
In many situations this minimization procedure is translated into the maximization of some measure of the ``size'' of the Fisher information matrix $I(\mu,\eta\vert\tau_a,\tau_b)$ of Eq.~\eqref{eq:fisher} (these can be eg. one of its norms or its determinant). This procedure can be seen effectively to be a minimization of the Cramer-Rao bound for an unbiased estimator of the target parameter.

Since in our application the maximum likelihood estimator Eq.~\eqref{eq:mle_estim} has a bias, we will minimize the mean squared error of $\mu_{mle}$ instead:
\begin{equation}
\label{eq:mse_cub}
\epsilon_M(\mu\vert\tau_a,\tau_b) = Var[\mu_{mle}] + B^2_\mu(\tau_a,\tau_b)\;.%\\
\end{equation}
One possible adaptive algorithm works by choosing, for any given iteration $i$, a new pair of time steps $(\tau^{i+1}_a,\tau^{i+1}_b)$ for the next rounds of $M$ measurements by minimizing Eq.~\eqref{eq:mse_cub} using the estimators $\widehat{\mu}_{mle}$ and $\widehat{\mu}_{mle}$ available at the current iteration $i$. The initial pair can be chosen randomly provided both time steps are small, in the results shown below we sample one of the two from a uniform distribution $U(0,0.1)$\footnote{The upperbound on $\tau$ here is somewhat arbitrary since the time step will get readjusted anyway.} while the second is chosen to minimize the following upper bound for the variance
\begin{equation}
Var\left[\mu_{mle}\right] \leq \frac{1}{M}\frac{\tau_a^6+\tau_b^6}{\tau_a^2\tau_b^2\left(\tau_a^2-\tau_b^2\right)^2}\;,
\end{equation}
keeping the first fixed.
This procedure ensures that the estimator for $\mu$ obtained from the new set of measurements have the smallest MSE possible and is thus rather efficient early on. As we collect more data and the variance of our estimator $\widehat{\mu}_{mle}$ gets reduced we should however reduce the contribution of the bias by reducing in magnitude the new pair of time steps. In order to incorporate this effect we obtain a new set $(\tau^{i+1}_a,\tau^{i+1}_b)$ of time steps by minimizing the expected variance of $\widehat{\mu}_{mle}$ after the new block of data is collected
\begin{equation}
\label{eq:mse_cub_it_ideal}
\epsilon_M^i(\mu\vert\tau_a,\tau_b) = Var[\mu_{mle}] + (i+1) B^2_\mu(\tau_a,\tau_b)
\end{equation}
which has the correct shot-noise scaling coming from $Var[\widehat{\mu}]\approx Var[\mu]/i$ as a function of the number of data blocks  $i$ collected so far. In practice we cannot evaluate the variance exactly and for the results presented in this work we found it sufficient to use the approximation
\begin{equation}
\label{eq:var_appr}
 \widetilde{Var}[\mu_{mle}]=\frac{4}{M}\frac{\tau_a^6\widetilde{P}_b\left(1-\widetilde{P}_b\right)+\tau_b^6\widetilde{P}_a\left(1-\widetilde{P}_a\right) }{\tau_a^2\tau_b^2\left(\tau_a^2-\tau_b^2\right)^2}
\end{equation}
where we have used
\begin{equation}
 \widetilde{P}_{a/b}  = \frac{1}{2}\left(1-\tau_{a/b}\widehat{\mu}_{mle}+\frac{\tau^3_{a/b}}{6}\widehat{\eta}_{mle}\right)
\end{equation}
to estimate the two probabilities $P_{a/b}$ (cf. Eq.~\eqref{eq:pair_prob}). Introducing an upperbound $B^u_\mu$ for the bias in Eq.~\eqref{eq:bias_cub}, the final cost function we use to find the new set of time steps is therefore
\begin{equation}
\label{eq:mse_cub_it}
\Delta_i = \left(\mu|\tau_a,\tau_b\right)\widetilde{Var}[\mu_{mle}]+(i+1) {B^u_\mu}^2(\tau_a,\tau_b)\;.
\end{equation}

The approximation of the variance Eq.~\eqref{eq:var_appr} is good in the limit of small time steps $\tau_{a/b}$ but this is not a problem since the presence of the bias term forces the optimal solutions to be numerically small automatically. In order to prevent numerical instabilities early on in the optimization we use a cost function that becomes extremely large whenever $\widetilde{P}_{a/b}\notin[0,1]$.

\section{The deuteron ground state}
\label{sec:deuteron}
The deuteron is the simplest nucleus present in nature. It is a bound state of a neutron and a proton in a state having total isospin $T = 0$, spin $S = 1$ and angular momentum-parity $J^\pi=1^+$. It has a small binding energy of approximately $2.2$ MeV.
The ground state of the deuteron has a non-zero quadrupole moment, originated by the mixing between s- and d-waves generated by pion-exchanges (see eg.~\cite{VanOrden2001} for a pedagogical introduction). A simple model for the system is to consider the 2-level system built from an s-wave orbital $\ket{\phi_S}$ and a d-wave one $\ket{\phi_D}$:
\begin{equation}
H = \begin{pmatrix}
\bra{\phi_S} H \ket{\phi_S} & \bra{\phi_S} H \ket{\phi_D}\\
\bra{\phi_D} H \ket{\phi_S} & \bra{\phi_D} H \ket{\phi_D}\\
\end{pmatrix}\;.
\end{equation}

Using the Argonne Av6' potential~\cite{Wiringa2002} we obtain (approximately~\footnote{The exact matrix elements with the Av6' potential are real, the integer representation used here is however a very good approximation and simplifies the discussion.}) the following Hamiltonian matrix:
\begin{equation}
\label{eq:det_ham}
H = \begin{pmatrix}
5 & -35\\
35 & 170\\
\end{pmatrix} = 87.5 \mathbb{1}-35 X + 82.5 Z\;.
\end{equation}
Large cancellations among different contributions produce a ground state energy orders of magnitude smaller than the norm:
\begin{equation*}
E_{gs} = -2.1174 \quad \|\overline{H_T}\|_1 = 117.5\quad R_O \approx 0.018  %\|\widehat{H_T}\|_2\approx86.617\;,
\end{equation*}
where, in analogy to the previous sections, we defined $H_T$ top be the traceless part of $H$ while $R_O$ is the ratio defined above in Eq.~\eqref{eq:ratio}. 

\begin{figure}
    \centering
    \includegraphics[scale=0.3]{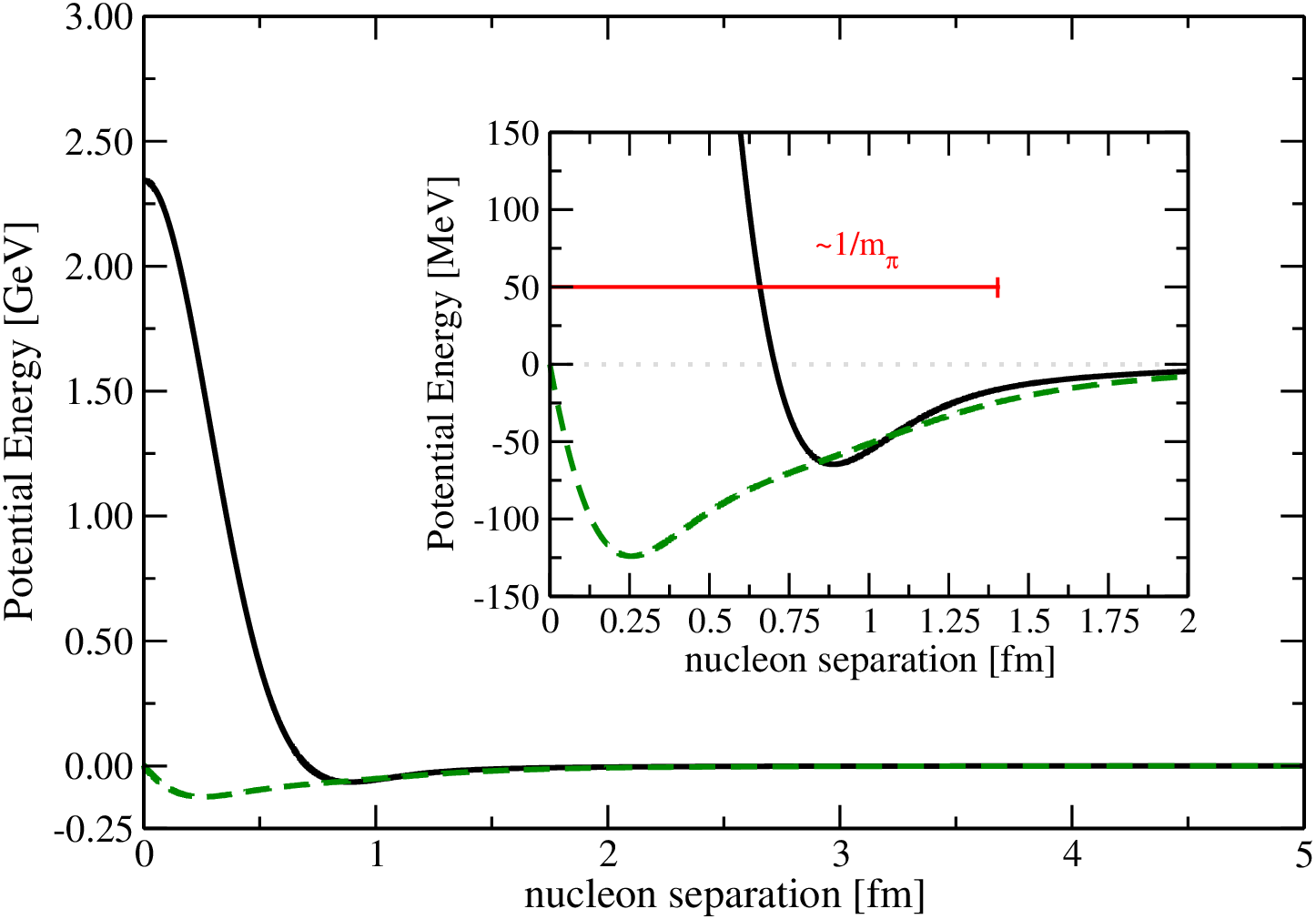}
    \caption{Central (black solid line) and tensor (green dashed line) contributions to the nuclear potential Argonne V6' used to obtain the deuteron Hamiltonian Eq.~\eqref{eq:det_ham}. In the inset the range of pion-exchange is also shown. The grey dotted line at $0$ energy is simply a guide to the eye.\label{fig:nuc_pot}}
\end{figure}

This large cancellation is a direct consequence of the hard-core nuclear repulsion that we can see in the central component of the interaction in Fig.~\ref{fig:nuc_pot} which introduces states with vary large energies in the many-body Hilbert space. This is a notorious problem which causes calculations on a finite basis to converge extremely slowly. General strategies to alleviate the issue have been developed in the past (eg. the Similarity Renormalization Group approach~\cite{Bogner2007,Hergert2007}) but they are usually accompanied by an increase in the degree on nonlocality of the Hamiltonian (see eg.~\cite{Anderson2010}) which in general will require a (possibly large) increase in the number of terms needed in expansions of the form Eq.~\eqref{eq:op_exp}.

Note however that, even when the detrimental effects of hard-core interactions are mitigated trough an effective theory like the one mentioned above, the requirement of ensuring basis-size convergence by performing multiple calculations with progressively larger basis sets will still lead to a possibly large mismatch between the ground state energy and the Hamiltonian norm. In that case this is due to fact that, as the basis size increases, the ground state energy will decrease at a much slower rate than the maximum eigenvalue (indeed $E_{gs}$ will reach a plateau for large basis while the highest eigenvalue will grow indefinitely). A general strategy to reduce the importance of this problem (like the sQPE scheme presented in this work) is thus welcome more generally.

Let's now start to discuss the performance of the Operator Averaging method of Sec.~\ref{sec:SM} on our model deuteron problem Eq.~\eqref{eq:det_ham}.
Using the estimate from Eq.~\eqref{eq:Ns_erel} we find that the number of measurement required for target relative accuracy $\epsilon_r$ is given by
\begin{equation}
\label{eq:mubound}
N_A(\epsilon_r)=\frac{1}{\left(R_O\epsilon_r\right)^2} \approx \frac{3079.4}{\epsilon_r^2} \quad\rightarrow\quad 3.1 \times 10^7\;,
\end{equation}
where the last limit holds for $\epsilon_r=1\%$. Even tough this estimate might not be very tight since we neglected the variances in Eq.~\eqref{eq:Ntot_bound} in order to derive Eq.~\eqref{eq:Ns_erel}, the fact that we are dealing with a simple one-qubit system and that we haven't considered yet the effect of errors, makes this requirement already alarming. In order to put this number in perspective, the IBM group~\cite{Kandala2017} estimated that $N_A\approx10^6$ measurements would be sufficient to reach chemical accuracy for a $6$ qubit model of $BeH_2$ with hundreds of Pauli terms in the Hamiltonian expansion of Eq.~\eqref{eq:op_exp}.

\begin{figure}
    \centering
    \includegraphics[scale=0.3]{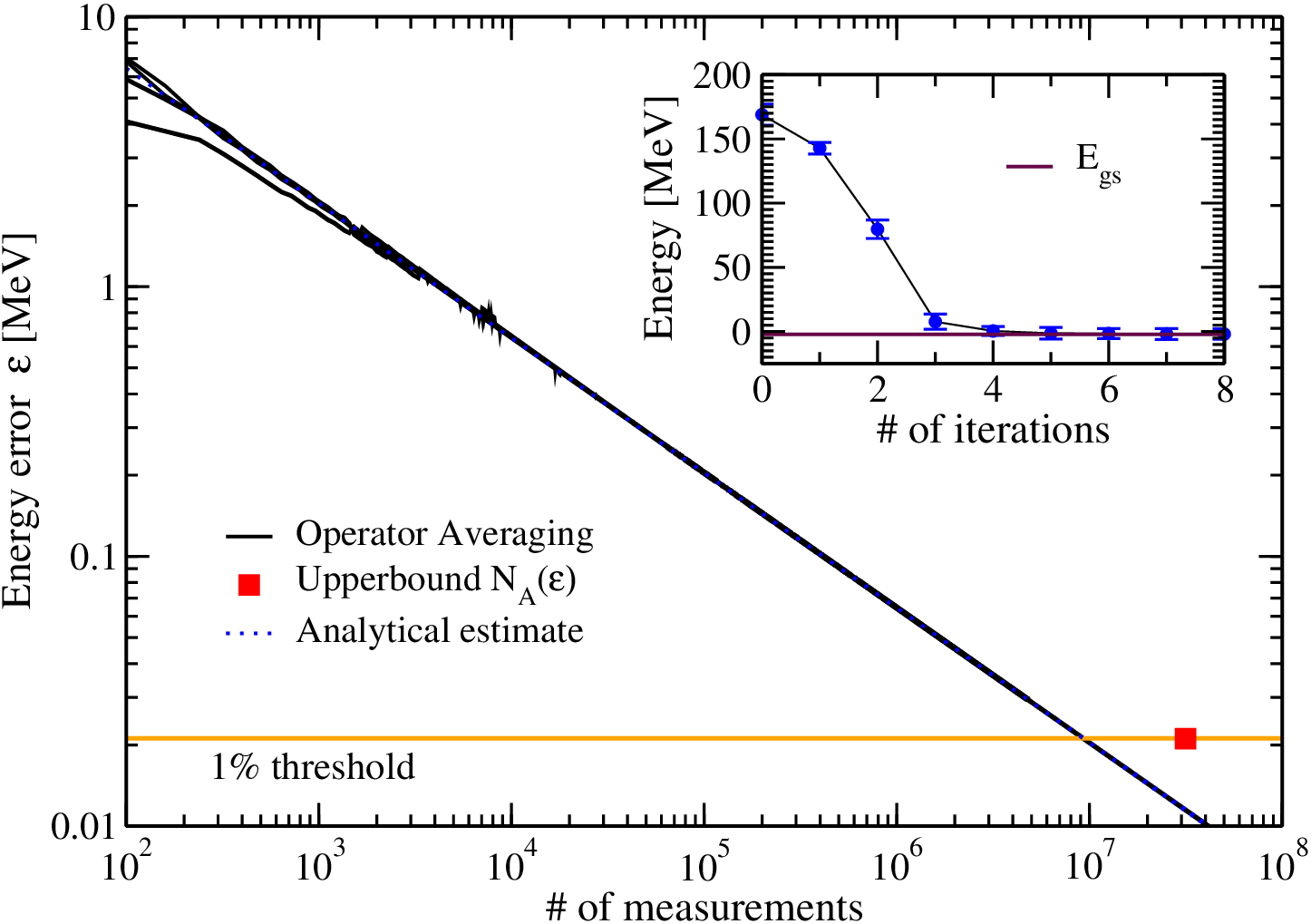}
    \caption{Results of numerical simulations of the algorithm explained in the text. Results correspond to 5 independent runs with different random number seeds. The blue dotted line corresponds to the expected asymptotic behavior from Eq.~\eqref{eq:err_oa} while the red dot correspond to the upperbound from Eq.\eqref{eq:mubound}. The inset shows the convergence of the VQE optimization using low resolution expectation values with errors $\approx5$ MeV (these were obtained using $N_{tot}=10^3$ measurements per function evaluation). The horizontal line indicates the value of the ground state energy $E_{gs}=-2.1174$ MeV.}
    \label{fig:sm}
\end{figure}

In Fig.~\ref{fig:sm} we show results for an ideal implementation (no noise apart from statistical fluctuations) of our one qubit model. As we can see the upperbound of Eq.~\eqref{eq:mubound} is only a factor of a few larger and we find that $\approx 9.3\times10^6$ measurements are needed in this ideal noiseless case. Before moving on to discuss the results we have obtained using sQPE, we want to point out that if our goal was only to optimize a variational state $\ket{\Phi(\vec{\theta})}$ using the energy expectation value (ie. we want to run VQE~\cite{Peruzzo2014}) then low accuracy results for the energy could be sufficient to get close to the optimum $\vec{\theta}_{min}$. To illustrate this we show in the inset the deuteron energy expectation value obtained using $N_{tot}=10^3$ as a function on the iteration in the minimization procedure (for these results we used a simple Nelder-Mead optimizer). Even though the error in the energy is $\gtrsim200\%$ the angle $\theta$ converges towards $\theta_{min}$ to within a few percent error in only a small number of iterations (note that for this simple model a single angle is sufficient to prepare the ground state). This striking difference is probably peculiar to simple models like our one-qubit deuteron, since the lack of excited states with low energy in the spectrum of the Hamiltonian gives rise to large gradients in the the variational energy $E(\theta)=\langle\Phi(\theta)\lvert H\rvert\Phi(\theta)\rangle$ and therefore to a relatively easy optimization. Where sQPE could be most useful in this case is for the final estimation of the energy, but in general for more complex systems low order sQPE could be advantageous also in the last stages of optimization where large statistical fluctuations could prevent to reach the minimum.

Let's now turn to discuss the sQPE method starting with the implementation of the controlled time evolution from Eq.~\eqref{eq:circuitcH}. In our simple two qubit situation the circuit for the controlled unitary can be constructed using only 2 CNOT gates (cf.~\cite{Barenco95}):
\begin{equation}
\Qcircuit @C=1em @R=.7em {
&\qw&\gate{\Phi(\theta_0)}&\ctrl{1}&\qw       &\ctrl{1}&\qw       &\qw\\
&\qw&\gate{R_A}           &\targ   &\gate{R_B}&\targ   &\gate{R_C}&\qw\\
}
\end{equation}
where $\Phi(\alpha)=diag\left(1,e^{i\alpha}\right)$ is a phase gate and the $R_A,R_B,R_C$ blocks are formed by appropriate single qubit rotations (see Appendix~\ref{app:circuit} for more details). %This circuit will be used in simulations of both the linear and the algorithm described in Sec.~\ref{sec:newmethod}.

\begin{figure}
    \centering
    \includegraphics[scale=0.3]{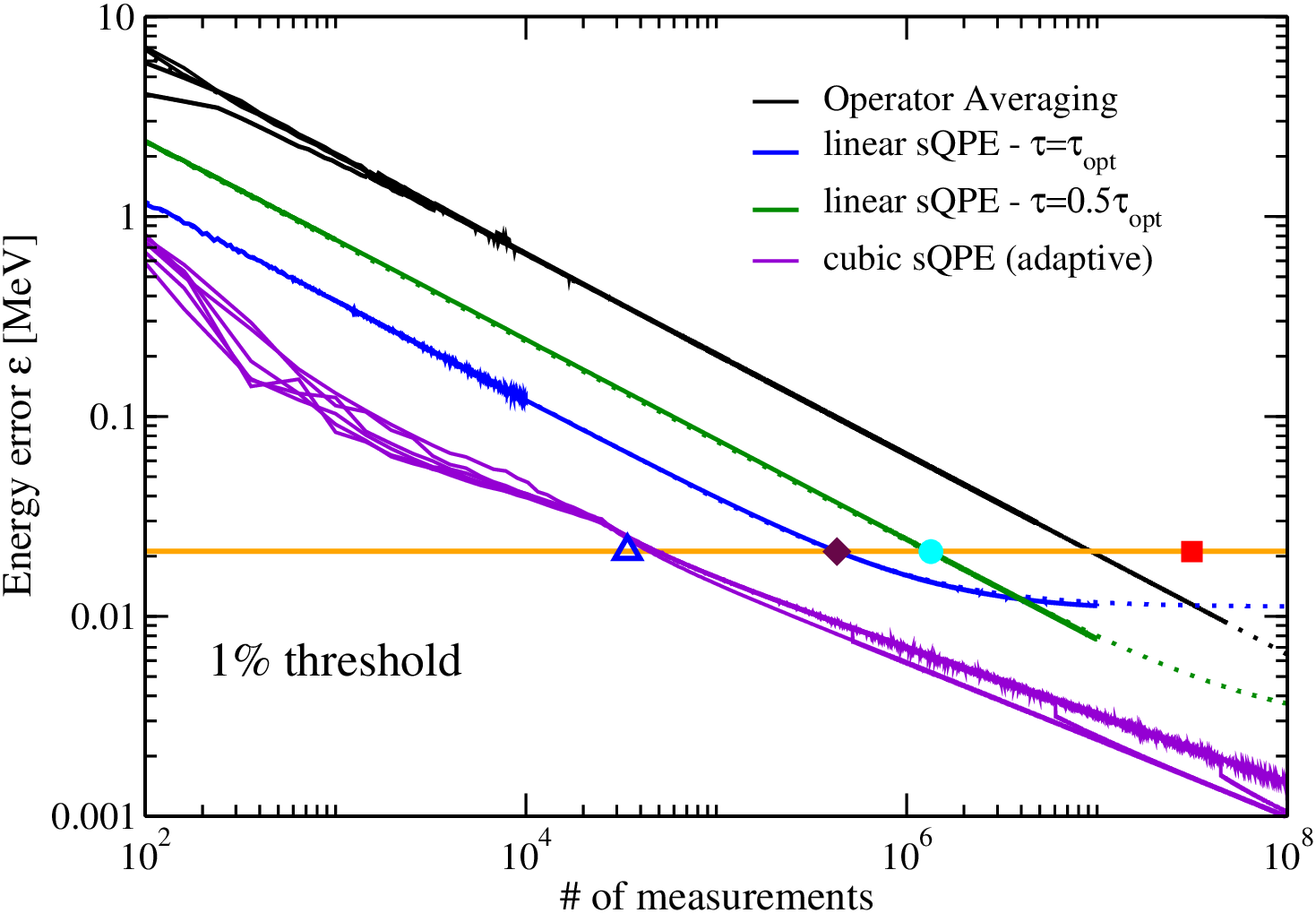}
    \caption{Final error in the estimator for the ground state energy of the deuteron obtained with the techniques discussed in this work. Dotted lines correspond to the analytical results presented in Fig.~\ref{figB}. See text for the meaning of the marked points.} 
    \label{fig:final_plot}
\end{figure}

In Fig.~\ref{fig:final_plot} we present the results obtained with both sQPE and operator averaging: the solid curves correspond to empirical results while the dotted lines correspond to the analytical estimates discussed in Sec.~\ref{sec:lin}. As for Fig.~\ref{figB}, the red square marks the location of the upperbound $N_A$ from Eq.~\eqref{eq:mubound}.
For the linear method, the optimal choice Eq.~\eqref{eq:tau_lin} is shown in blue, while in green we present the results obtained using a more conservative value $\tau=\tau_{opt}/2$ (cf. Eq.~\eqref{eq:atau_lin} and the discussion following it). In both cases we see that, in agreement with the results presented in Fig.~\ref{figB}, the linear algorithm requires about an order of magnitude less measurements than operator averaging. In addition to this we see that the estimated number of measurements $N_{sQPE}$ are in very good agreement with the empirical results: the maroon diamond indicates the upperbound $N_{sQPE}(1)$ for linear sQPE Eq.~\eqref{eq:lin_1perc} while the cyan circle corresponds to the worse bound obtained trough Eq.~\eqref{eq:lin_app}. 

As we discussed in Sec.~\ref{sec:lin} the speedup offered by the linear method is very sensitive to the particular choice of time step used and by employing the adaptive strategy described in Sec.~\ref{sec:cub} to find the time steps $(\tau_a,\tau_b)$ the cubic algorithm partially overcomes this problem. In order to implement the algorithm we use the same circuit described for the linear method: we just run it twice for the two time steps separately. The results obtained from $6$ different runs are presented in Fig.~\ref{fig:final_plot} as purple lines, in all cases we update the time step pair every block of $N_b=40$ measurements. In addition, the blue triangle is twice the expected number of measurements $N_{sQPE}(2)$ obtained from the general expression Eq.~\eqref{eq:NT}
\begin{equation}
N_{sQPE}(2,\epsilon_r) = \frac{f(2)}{\epsilon^{5/2}}\sqrt{\left\lvert\langle H^5\rangle\right\rvert}\approx 1.7\times10^4\;,
\end{equation}
where the factor of $2$ is introduced to account for the fact that in our adaptive scheme we are actually estimating two expectation values: $\langle O\rangle$ and $m_1$.
The optimal pair of time steps $(\tau_a,\tau_b)$ obtained during the execution of the algorithm fluctuate around the value $(0.15,0.3)$ which is not very far from the optimal time step $\tau_{opt}\approx0.4$ found from Eq.~\eqref{eq:time_step}. The spread of results at large measurement count is possibly a signature that the optimization of Eq.~\eqref{eq:mse_cub_it} gets stuck in local minima, we plan to investigate this further in future work.

As explained in Sec.~\ref{sec:cub}, the cubic algorithm needs a good approximation of the bias term $B_\mu(\tau_a,\tau_b)$ in  Eq.~\eqref{eq:bias_cub} as this enters directly the cost function Eq.~\eqref{eq:mse_cub_it} used to determine the optimal time steps. The results presented above where obtained using the following ansatz
\begin{equation}
\label{eq:ba1}
 B^u_\mu(\tau_a,\tau_b)= \frac{ \lvert \widehat{\mu}_{mle} \widehat{\eta}_{mle} \rvert}{120}\tau_a^2\tau_b^2\frac{\tau_a^2+\tau_b^2}{\lvert \tau_a^2-\tau_b^2 \rvert}\equiv B_{A1}\;,
\end{equation}
where $\widehat{\mu}_{mle}$ and $\widehat{\eta}_{mle}$ are the current best estimators for the distribution parameters $(\mu,\eta)$. This form reduces to the correct one in the eigenvalue estimation limit relevant here where $\ket{\Psi}$ is an eigenstate of $O$.
We want now to present results showing the weak sensitivity of the cubic algorithm to the specific choice of the estimator for the bias, in particular we will use two additional estimators: the exact one from Eq.~\eqref{eq:bias_cub_e}
\begin{equation}
B_E =B^u_\mu(\tau_a,\tau_b) = \mathbb{E}\left[\mu_{mle} - \mu\right],
\end{equation}
which in practical situations we won't have access to, and a different variant of the estimator $B_{A1}$ above defined as
\begin{equation}
\label{eq:ba2}
 B_{A2}= \frac{ \lvert \widehat{\mu}_{mle} \widehat{\eta}_{mle} \rvert}{120}\tau_a^2\tau_b^2\frac{\max{\left[\tau_a^2,\tau_b^2\right]}}{\lvert \tau_a^2-\tau_b^2 \rvert}\;.
\end{equation}
This estimator is a tighter bound that can be obtained from  Eq.~\eqref{eq:r5} using the additional condition
\begin{equation}
\max{\left[\tau_a^2,\tau_b^2\right]} < \frac{\pi}{\|\overline{O}\|_1} \;.
\end{equation}

\begin{figure}
    \centering
    \includegraphics[scale=0.3]{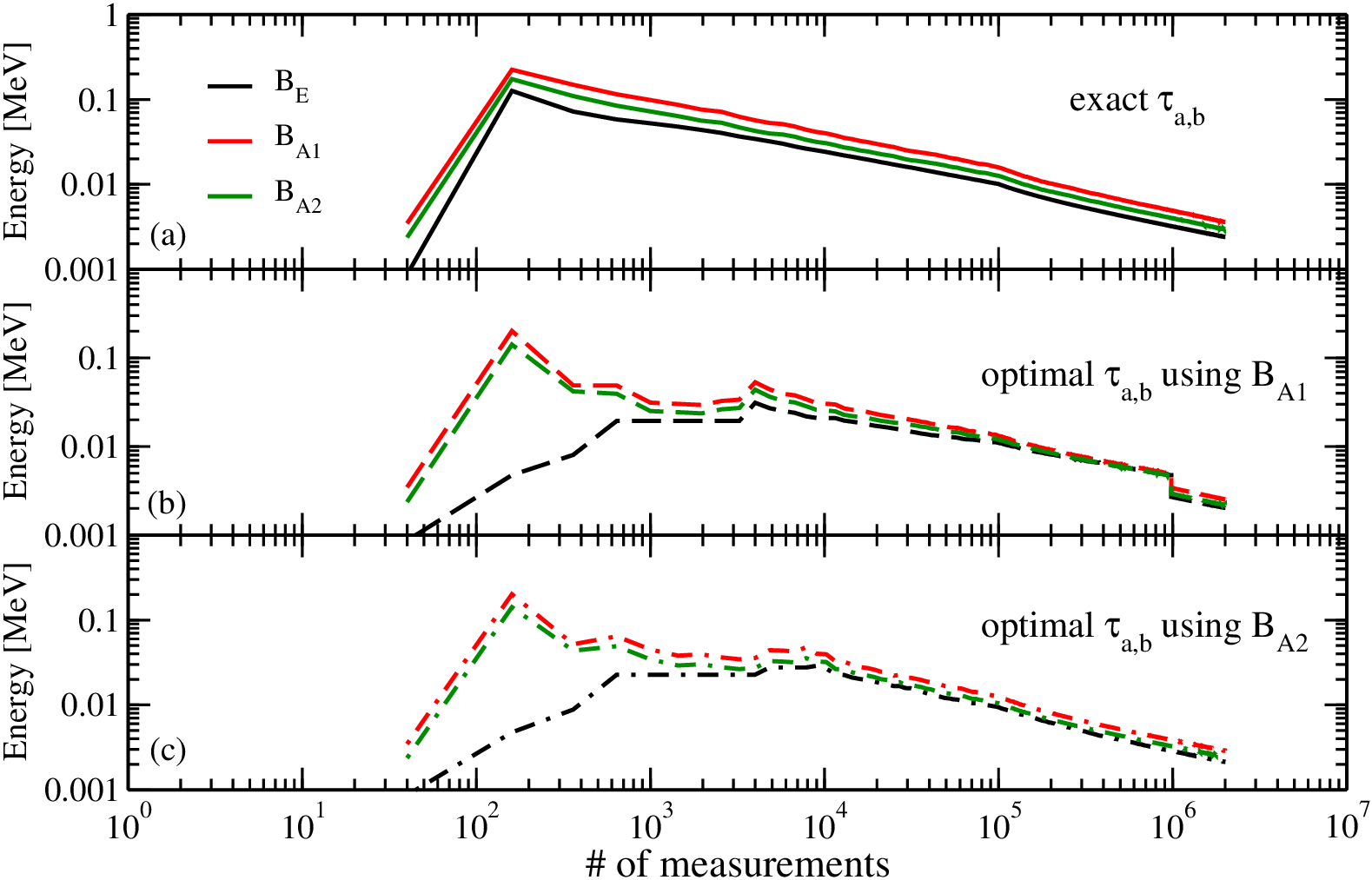}
    \caption{Different estimators of the energy bias for a cubic sQPE calculation of the deuteron ground state energy (see text for the definitions). The top panel shows results for an idealized case while panel $(b)$ and $(c)$ are obtained using approximate optimizations employing the estimators $B_{A1}$ and $B_{A2}$ (cf. Eq.~\eqref{eq:ba1} and Eq.~\eqref{eq:ba2}) }
    \label{fig:biasA}
\end{figure}

In Fig.~\ref{fig:biasA} we show how these estimators evolve as the algorithm proceeds in three different situations. The top panel shows the ideal situation where the optimal pair $(\tau_a,\tau_b)$ for the next step is obtained using the exact scaled mean squared error (cf. Eq.~\eqref{eq:mse_cub_it_ideal})
\begin{equation}
\epsilon_M^i(\mu\vert\tau_a,\tau_b) = Var[\mu_{mle}] + (i+1) B^2_\mu(\tau_a,\tau_b)\;,
\end{equation}
together with the exact bias $B_E$. Even though in practical situation we won't be able to run the cubic algorithm this way, these results provide a ceiling for the performance of approximate algorithms while at the same time show clearly the source of the advantage that is achieved with sQPE: initially the time steps are raised to relatively large values in order to reduce the shot noise limited variance term in the equation above at the expense of a larger bias term. This allows to quickly reduce the error in the expectation value early on when the dominant contribution are statistical fluctuations, as the accuracy increases the importance of the bias term grows and the adaptive algorithm starts to reduce the magnitude of the time steps in order to keep $B_\mu$ under control. Furthermore we see that the three bias estimators follow each other rather closely.

The other two panels instead correspond to results obtained using the approximate cost function $\Delta_i$ from Eq.~\eqref{eq:mse_cub_it} with either the $B_{A1}$ ansatz used also in Fig.~\ref{fig:final_plot} (central panel) or the $B_{A2}$ ansatz described above (bottom panel). In both situations we recover the same qualitative behavior seen in the ideal case: the bias gets initially increased and then reduced gradually as the accuracy improves. The main difference with the results of the top panel is the lower efficiency obtained in the first stage of this procedure where the actual exact bias remains much smaller than it could have been for the first few hundred measurements (note that as above we update the time-steps every $N_b=40$ measurements) but the discrepancy quickly vanishes later when the bias becomes the limiting factor.

\begin{figure}
    \centering
    \includegraphics[scale=0.3]{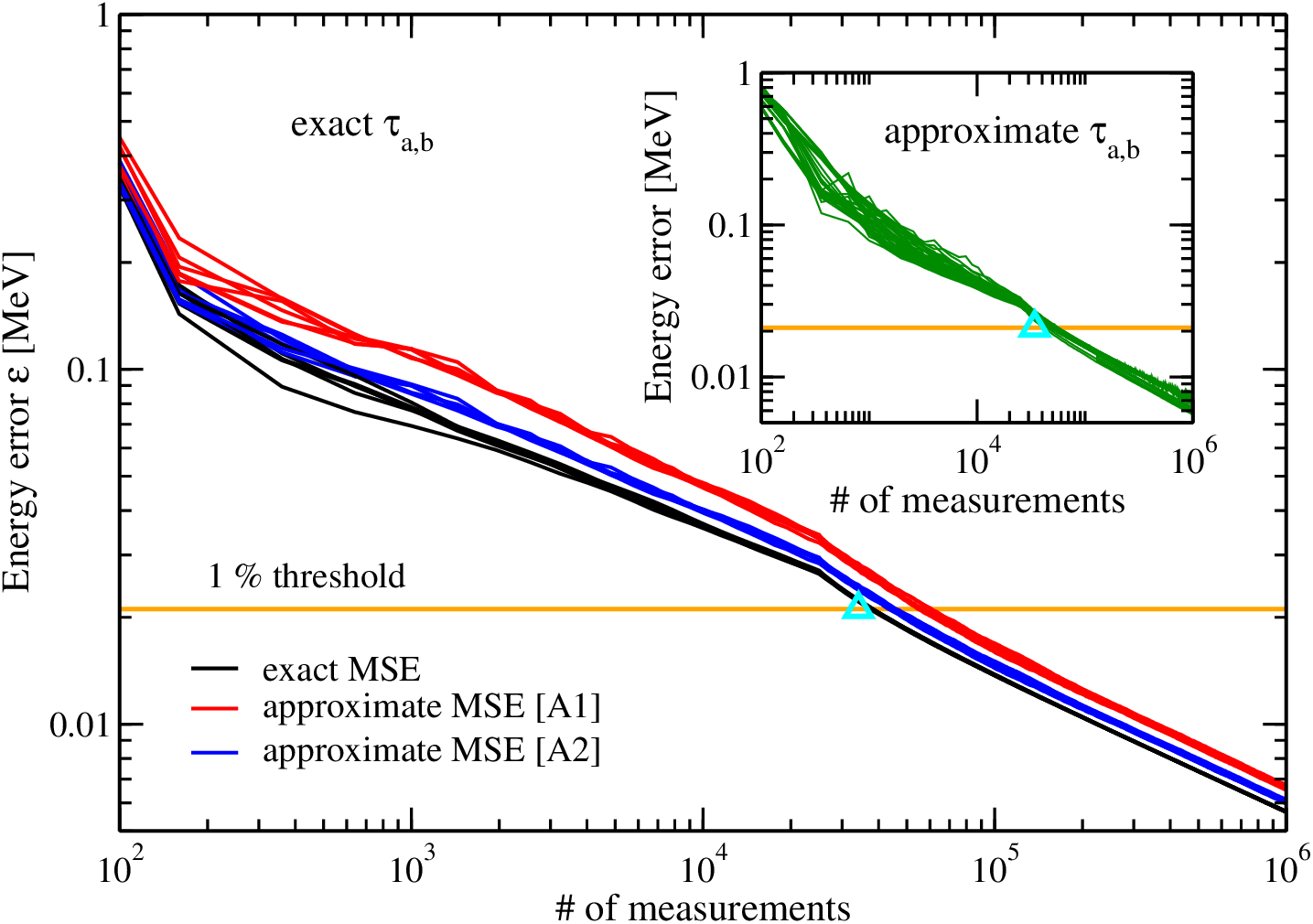}
    \caption{Different estimators for the final error in the deuteron ground state energy form ideal runs where the time-steps are optimized exactly using Eq.~\eqref{eq:mse_cub_it_ideal}. The cyan up triangle is the same as the blue triangle in Fig.~\ref{fig:final_plot}. The inset show results obtained using the approximate cost function Eq.~\eqref{eq:mse_cub_it}.}
    \label{fig:biasB}
\end{figure}

The see the impact of these approximations on the final convergence of the expectation value we show in Fig.~\ref{fig:biasB} the results obtained by using the exact pair of time steps $\tau_{a/b}$ obtained as before using Eq.~\eqref{eq:mse_cub_it_ideal} but different approximations to the final mean squared error. The reason this is important is that we need to estimate the bias in order to provide an estimate for the final error of our estimated expectation value. The results in the main panel show the apparent reduced efficiency that is a result of using a bigger bias than the exact one (in this case we used $B_{A1}$ and $B_{A2}$ as above). The cyan up triangle is the same as the blue triangle in Fig.~\ref{fig:final_plot} and indicates an optimistic expectation on the efficiency of the cubic algorithm. Somewhat not surprisingly we achieve that estimate only in the ideal exact case shown as black lines in the main panel of Fig.~\ref{fig:biasB} while in the worst case (corresponding to the red curves) we need to perform as much as $\approx50\%$ more measurements.
Interestingly when statistical fluctuations in the estimation of the variance are included these differences mostly disappear and there is no clear preference for different choices of the bias, we can see this from the results shown in the inset of Fig.~\ref{fig:biasB} where the time steps were estimated using the approximate cost function $\Delta^i$ from Eq.~\eqref{eq:mse_cub_it}. 

These results are encouraging as they show that, even if a tight upperbound for the bias is helpful for the algorithm, an approximate expression can work very well at least for the special case of eigenvalue estimation.
Evaluating upperbounds becomes more important in the general case and we leave the discussion for Sec.~\ref{sec:pcond} while we now turn to the problem of accounting for the presence of noise in the quantum device.

\section{Effect of noise}
\label{sec:noise}
As we have mentioned in the introduction, noise will be unavoidable for near term quantum devices and it is therefore critical for algorithms to provide robustness against noise if we want to deploy them on a non fault-tolerant quantum computer.
Since the methodology we propose goes against this trend in that we are trading classical resources (the number of experimental trials) with quantum ones (one more qubit for the ancilla and more gates), we need to provide supporting evidence that our method shows advantages even in the presence of noise and is thus practical. 

The importance of this assessment is critical as there are known cases where the advantage of a quantum algorithm can be drastically reduced by the presence of even small noise sources (see eg.~\cite{Shenvi2003,PE2010,HL2012,Frowis2018}).

We start by showing how, in situations where condition Eq.~\eqref{eq:cond} is valid by a large margin (as in the deuteron model discussed here), a substantial increase in classical resources is required to minimize the effect of measurement noise for the Operator Averaging method of Sec.~\ref{sec:SM} while with sQPE this problem can be mitigated. In the last part we provide a more general argument in support of our measurement strategy in situations where a clean ancilla is available in the spirit of the D1QC model~\cite{Knill98}. 
 
\subsection{Measurement noise}
\begin{figure}
    \centering
    \includegraphics[scale=0.3]{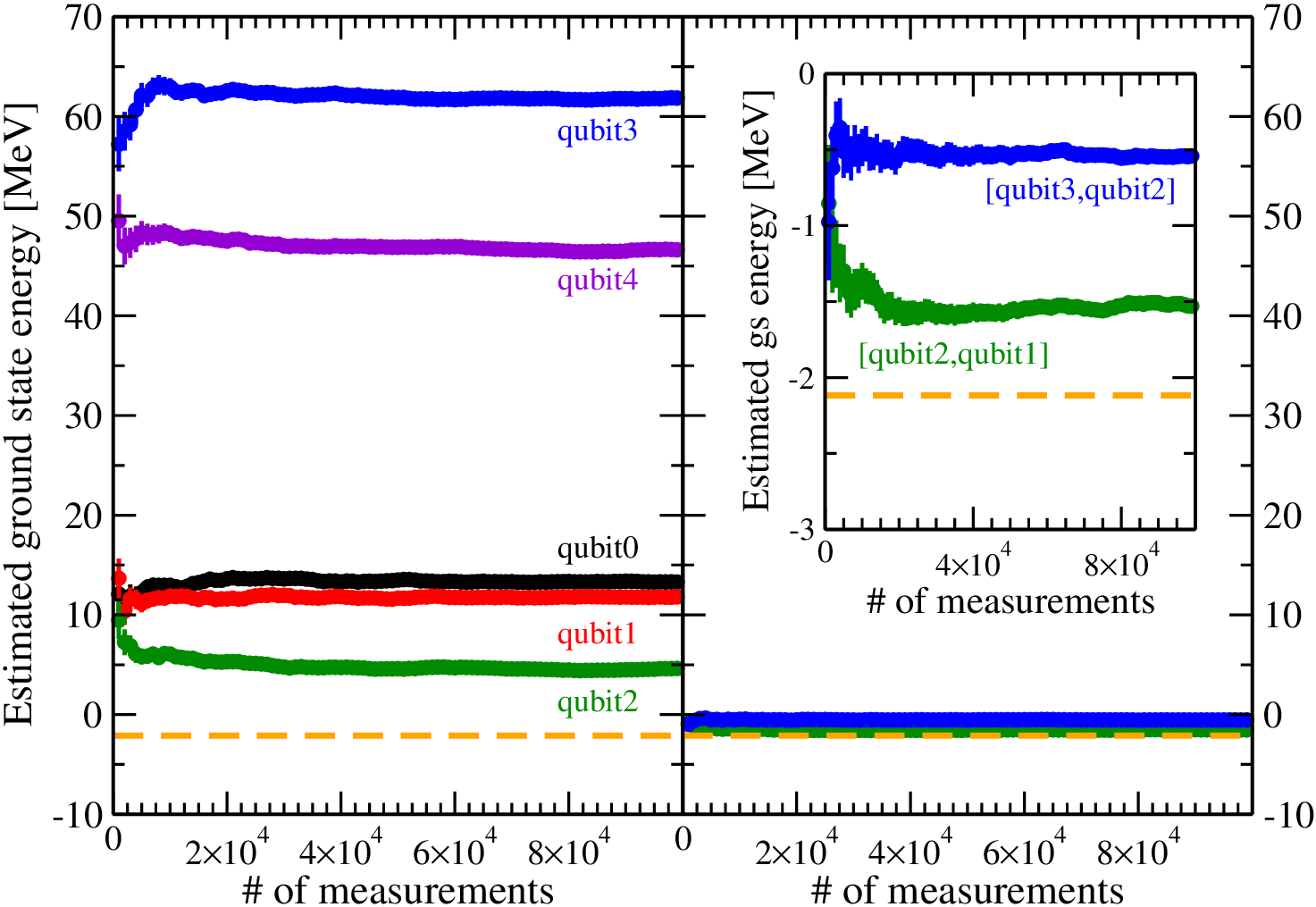}
    \caption{Results for the deuteron ground state energy obtained using the emulated 'ibmqx4' quantum computer without error mitigation. The left panel shows results obtained using Operator Averaging on each qubit while the right panel shows the performance of the linear sQPE algorithm using pairs $[qubit2,qubit1]$ (green data points) and $[qubit3,qubit2]$ (blue data points) as ancilla and system qubit respectively.
    \label{fig:noice_cmp}}
\end{figure}

Assignment errors in the measurement device used for qubit read out are an important source of bias that needs to be accounted for properly in order to obtain meaningful results. In order to illustrate the problem we show in Fig.~\ref{fig:noice_cmp} results obtained by executing the deuteron problem described in the previous section on a emulated version of the IBM 5-qubit machine 'ibmqx4' using the Qiskit software package~\cite{Qiskit} (see Tab.~\ref{tab:noice_par} for details).

\begin{table}[]
\begin{tabular}{c|l|l}
qubit & Rotation err & Readout err \\ \hline
0 & 0.0019 & 0.0865 \\
1 & 0.0024 & 0.08 \\
2 & 0.0024 & 0.0382 \\
3 & 0.0027 & 0.3567 \\
4 & 0.0036 & 0.2715
\end{tabular}
\caption{Rotation (U3) and readout errors for IBM's 5-qubit machine 'ibmqx4' on May 8 2019. The error on the CNOT gate on the pairs $[qubit2,qubit1]$ and $[qubit3,qubit2]$ is $4.88\%$ and $6.68\%$ respectively.}
\label{tab:noice_par}
\end{table}

Despite the fact that we are not correcting for any source of errors in these results, the linear sQPE (with optimal time $\tau$) presented on the right seems to provide considerably higher quality results despite the much increased circuit depth. In the following we provide an argument to explain the observed results.

\begin{figure}
    \centering
    \includegraphics[scale=0.3]{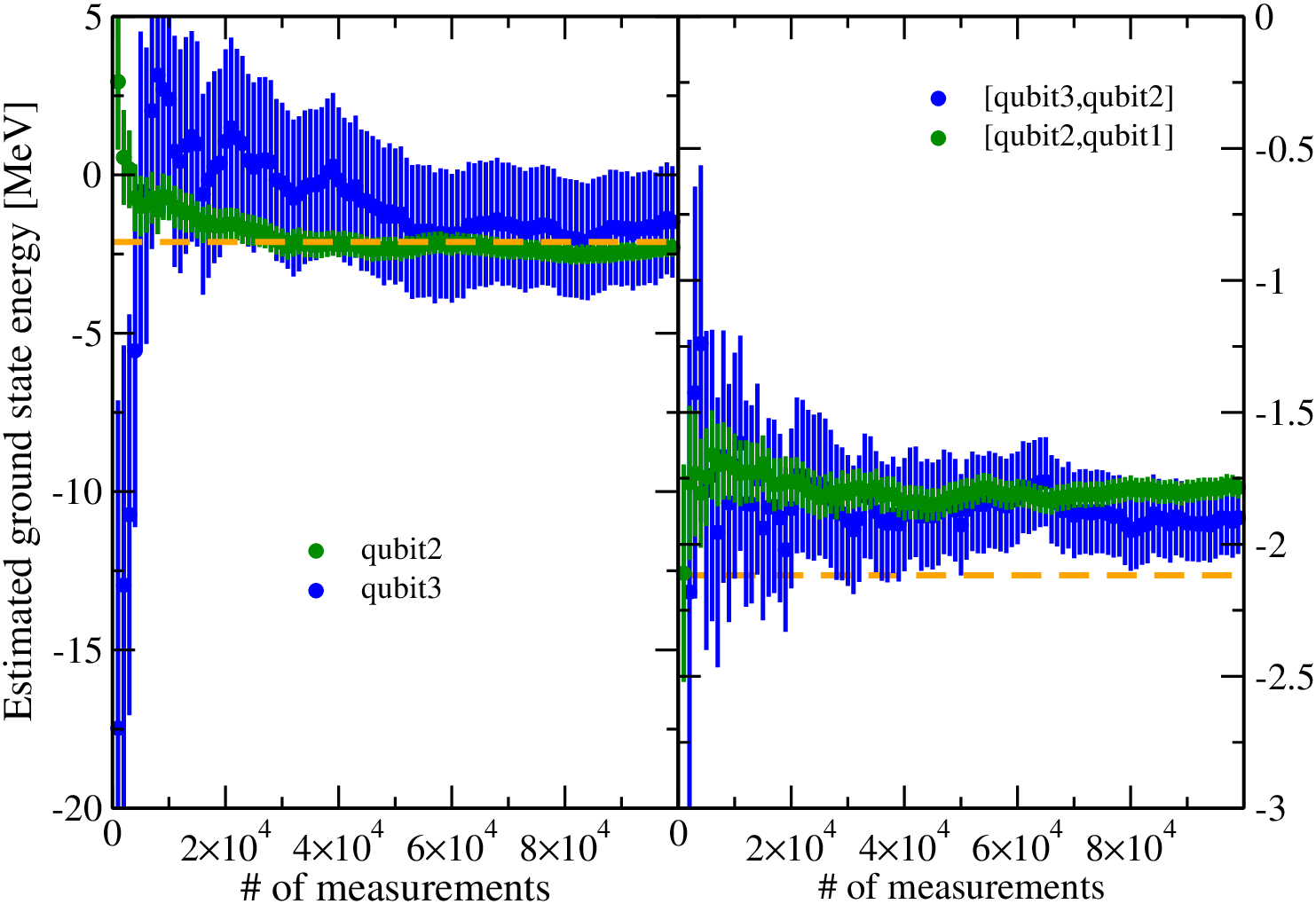}
    \caption{Same as in Fig.~\ref{fig:noice_cmp} but using the error mitigation strategy described in the text for the measurement noise. \label{fig:noice_fix}}
\end{figure}

Here we will use an extremely simplified model for these errors that nevertheless captures their essential features, this is achieved by replacing the projectors $\Pi_0$, $\Pi_1$ on the states $\ket{0}$, $\ket{1}$ of a qubit with the following ones
\begin{equation}
\label{eq:ro_noice}
\begin{split}
\widetilde{\Pi_0} &= \left(1-p\right) \Pi_0 + p \Pi_1\\
\widetilde{\Pi_1} &= \left(1-p\right) \Pi_1 + p \Pi_0
\end{split}
\end{equation}
where $0< p<1$. This model can be justified in the limit where assignment errors are both qubit independent and symmetric with respect to the interchange $\ket{0} \leftrightarrow \ket{1}$ and is sufficient for our purpose (see eg. Supplemental Material of ~\cite{Kandala2017} for details on a more accurate model). Using Eq.~\eqref{eq:ro_noice} we find that the noisy expectation value of some one-qubit Pauli operator $\langle\widetilde{P_\sigma}\rangle$ is related to the noise-free value by the relation:
\begin{equation}
\label{eq:ro_error_corr}
\langle\widetilde{P_\sigma}\rangle = \left(1-2p\right) \langle P_\sigma\rangle\;,
\end{equation}
which can be easily inverted to estimate $\langle P_\sigma\rangle$ from $\langle\widetilde{P_\sigma}\rangle$. Despite it's simplicity this model is sufficient to completely account for the error afflicting the OA results of Fig.~\ref{fig:noice_cmp} as we can see from the error mitigated results presented in Fig.~\ref{fig:noice_fix}. Note that the linear sQPE energies in the right panel are still biased due mostly to the noise introduced by using the CNOT gates, the mitigation of which is beyond the scope of our discussion here (note however that mitigation techniques~\cite{Temme2017,Endo2018} will be required also for OA for larger target systems).

Generalizations of Eq.~\eqref{eq:ro_error_corr} to expectation values of multi-qubit Pauli operators can also be obtained~\cite{Kandala2017}, but we will limit our discussion to the one-qubit case relevant to our deuteron calculations and to sQPE more generally. In fact, an important feature of sQPE is that the output of the algorithm is obtained through measurements on a single qubit thus avoiding the problem of exponentially reduction of signal to noise ratio as a function of the number of qubits involved in the measurement of individual terms in the expansion Eq.~\eqref{eq:op_exp} (see eg.~\cite{Kandala2017}).

For a generic one-qubit observable
 \begin{equation}
O = \alpha_0 + \sum_{i=1}^3 \beta_i P_i \quad\vec{P}=\left(X,Y,Z\right)\;,
 \end{equation}
we can now use Eq.~\eqref{eq:ro_error_corr} to estimate the noise free expectation value
\begin{equation}
\langle \widehat{O}\rangle = \alpha_0+\sum_{i=1}^3 \frac{\beta_i }{1-2\widehat{p}} \langle \widetilde{P_i} \rangle\;,
\end{equation}
where $\widehat{p}$ is a finite sample estimator estimator, with variance $\delta p$, of the error probability $p$. The variance of this estimator can be approximated as
\begin{equation}
\label{eq:var_ro}
Var[\widehat{O}] = \frac{Var[\widetilde{O}]}{(1-2\widehat{p})^2}+ V_{R}[O];,
\end{equation}
with
\begin{equation}
V_{R} = \frac{4\delta p^2}{(1-2\widehat{p})^2} \sum_{i=1}^3\frac{|\beta_i|^2}{(1-2\widehat{p})^2}\langle\widetilde{P_i}\rangle^2\;,
\end{equation}
where we used a linear expansion to propagate the error (ie. we used $Var[f(x)]\approx f'(x)^2Var[x]$). The second error term in Eq.~\eqref{eq:var_ro} comes from the uncertainty in the determination of the error parameter $p$ and provides a noise floor that we need to minimize in order to achieve good accuracies. Assuming the sample estimator $\widehat{p}$ was obtained from $N_C$ calibration measurements, we can bound the contribution of this background as
\begin{equation}
V_{R} \leq 4 \frac{\|\overline{O_T}\|^2_2}{(1-2\widehat{p})^4}\delta p^2=4 \frac{\|\overline{O_T}\|^2_2}{(1-2\widehat{p})^4} \frac{\widehat{p}\left(1-\widehat{p}\right)}{N_C}\;,
\end{equation}
where in the first step we used the decomposition Eq.~\eqref{eq:op_exp}.

The case of the sQPE algorithm is simpler because we have always to deal with a single qubit to be measured. Using the same correction scheme employed above, assuming again that the higher order coefficients $m_k$ are known, we find
\begin{equation}
\langle \widehat{O_K}(\tau)\rangle = -\frac{1}{\tau}\frac{\langle Z \rangle }{1-2\widehat{p}}+\sum_{k=1}^{K-1}\tau^{2k}\frac{(-1)^k m_k}{(2k+1)!}\;,
\end{equation}
with variance
\begin{equation}
\label{eq:var_ro_sqpe}
Var[\widehat{O_K}(\tau)] = \frac{Var[\widetilde{O}_K(\tau)]}{(1-2\widehat{p})^2}+ V_{RK}[O_K(\tau)];,
\end{equation}
and
\begin{equation}
\begin{split}
V_{RK}[O_K(\tau)] &= \frac{4}{\tau^2}\frac{\delta p^2}{(1-2\widehat{p})^2}\frac{\langle Z \rangle^2}{(1-2\widehat{p})^2}\\
&\leq\frac{4}{\tau^2} \frac{1}{(1-2\widehat{p})^4} \frac{\widehat{p}\left(1-\widehat{p}\right)}{N_C} \;.
\end{split}
\end{equation}
It is then clear that the sQPE algorithm will reduce the importance of measurement noise in the same situations where it provides an advantage in the noise free case, namely whenever $ \|\overline{O_T}\| \gg \frac{1}{\tau^2}$ with $\tau$ the estimated optimal time step for a particular problem.

\begin{figure}[h]
    \centering
    \includegraphics[scale=0.3]{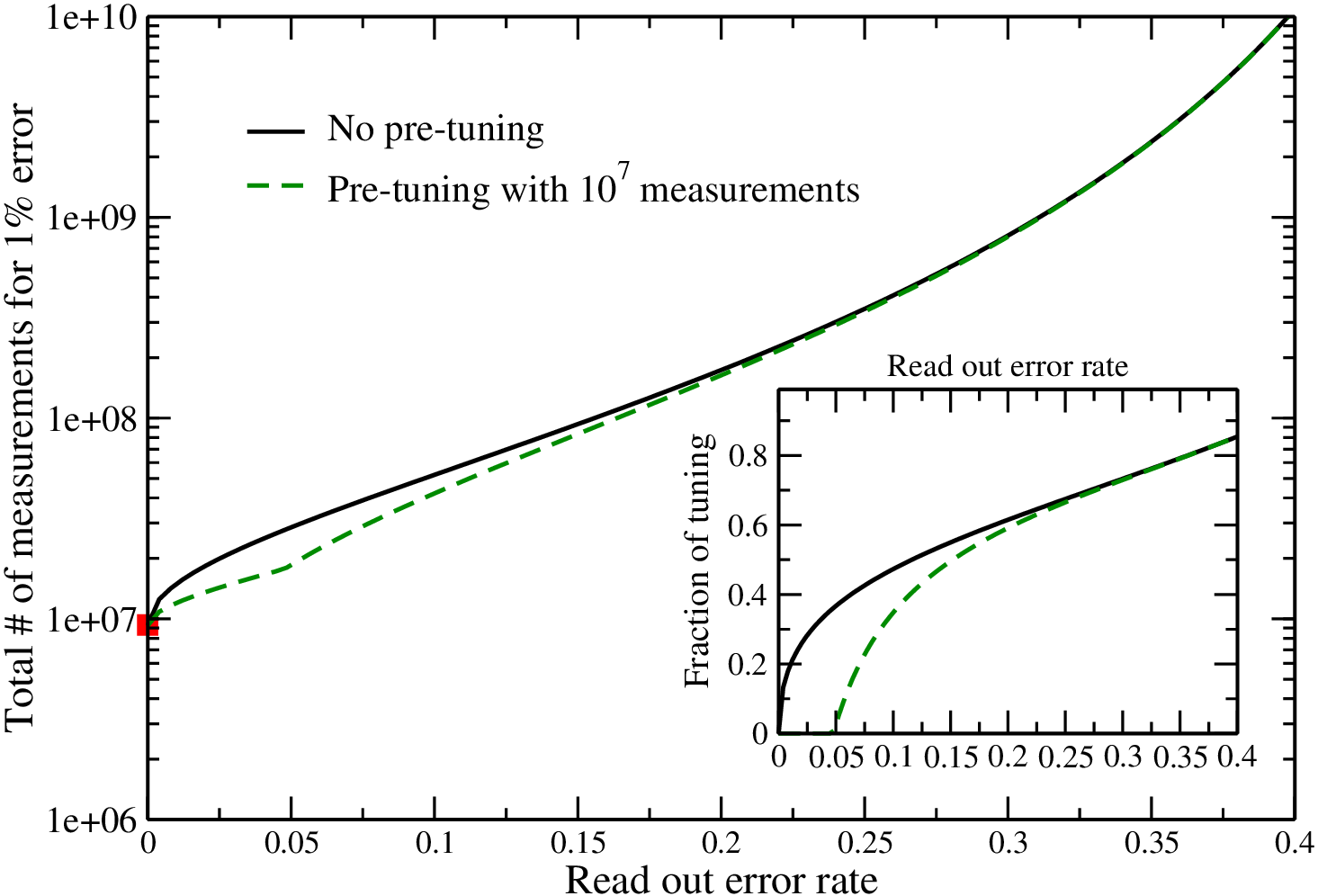}
    \caption{Estimated number of measurement required to reproduce the ground-state energy of the deuteron to $\epsilon_r =  1\%$ accuracy as a function of the read-out error rate. Continuous lines show are the total number ($N_{tot}=2N+N_C$) while the dashed lines show the fraction of the total measurements that has to be dedicated to characterize the readout noise.
    \label{fig:ro_noice}}
\end{figure}

In order to assess the practical impact of this error term for the deuteron calculation of Sec.~\ref{sec:deuteron} we have numerically minimized the total number of measurements $N_{tot}=2N+N_C$ needed to achieve $\epsilon_r=1\%$ ($N$ for each of the Pauli terms and $N_C$ to estimate $\widehat{p}$) as a function of the error parameter $p$ using directly Eq.~\eqref{eq:var_ro}.
In Fig.~\ref{fig:ro_noice} we show the results of this study: the full black line is the minimal value of $N_{tot}$ needed to reach a target relative error $\epsilon_r=1\%$, while the green dashed line corresponds to the situation where we have performed a calibration using $N_C^0=10^7$ initial measurements to estimate the error rate $p$ of the machine.

The inset shows the ratio of tuning measurements needed to achieve the target accuracy as a function of the error rate $p$ in both situations. We see that small values of the error rate $p\lesssim5\%$ can be dealt with with relative ease using data obtained from previous calibrations, while more substantial efforts are needed for more noisy qubits. Using $qubit3$ with its large error $p\gtrsim35\%$ would require two orders of magnitude more measurements than in the noiseless case (shown as the red square in Fig.~\ref{fig:ro_noice}) out of which more than $\approx80\%$ would be needed for characterizing the noise. Fortunately noise levels so high are not common on modern machine, with typical values of a few percent in the superconducting circuit case, but the fact that this very simple source of error is capable of completely swarming the results of a simple single qubit calculation provides another motivation to explore the use of the sQPE scheme on near term architectures.

\subsection{General advantage of ancilla-based schemes}

The purpose of this section is to show that the ancilla-based construction of sQPE (cf. Eq.~\eqref{eq:circuitcH}) can be advantageous in general when we want to estimate the value of an expectation value $\langle O\rangle$ in presence of depolarizing noise (see eg.~\cite{NandC_book}) in the quantum device. For a realistic advantage to be found we will assume we have either a clean ancilla (meaning completely error free) or at least a qubit subject to a well characterized noise channel and with high measurement fidelity.

Let's start by considering a slight generalization of the sQPE circuit Eq.~\eqref{eq:circuitcH} where we leave unspecified the state of the ancilla before we apply the controlled unitary
\begin{equation}
\label{eq:circ_ext}
\Qcircuit @C=1em @R=1em {
\lstick{\ket{\phi}} & \ctrl{1} &  \qw \\
\lstick{\ket{\Psi}}   & \gate{U_\tau} &  \qw}
\quad=\quad\quad\quad\Qcircuit @C=1em @R=1em {
\lstick{\ket{0}} & \gate{W}&\ctrl{1} &  \qw \\
\lstick{\ket{\Psi}} &\qw  & \gate{U_\tau} &  \qw}\;,
\end{equation}
and in the second circuit we made explicit the presence of a new rotation matrix $W$ which prepares $\ket{\phi}$ starting from $\ket{0}$. The action of the time evolution unitary $U_\tau$ on the target state $\ket{\Psi}$ can be conveniently expressed as
\begin{equation}
U_\tau \ket{\Psi} = \kappa\ket{\Psi}+\nu \ket{\Psi^\perp}%\quad\text{with}\quad\langle\Phi\vert\Phi^\perp\rangle=0
\end{equation}
with $\kappa,\nu\in\mathbb{C}$ with
\begin{equation}
\langle\Psi\vert\Psi^\perp\rangle=0\quad\text{and}\quad\lvert\kappa\rvert^2+\lvert\nu\rvert^2=1\;.
\end{equation}
In other words, the total Hilbert space explored with circuits of type Eq.~\eqref{eq:circ_ext} is only $4$-dimensional: we have $\mathbb{C}^2$ for the ancilla and the linear span of $\ket{\Psi}$ and $\ket{\Psi^\perp}$ for the target system. 
At  this point, a measurement of the $y$-polarization of the ancilla after the circuit in Eq.~\eqref{eq:circ_ext} will reveal the wanted quantity (cf. Sec.~\ref{sec:newmethod})
\begin{equation}
\langle Y\rangle_a = \mathcal{I}\left[U_\tau\rvert\Psi\rangle\langle\Psi\lvert\right] =\mathcal{I}\left[\kappa\right]\equiv\kappa_I
\end{equation}
from which we can extract the expectation value as discussed above.

If we trace out the system qubits, the circuit in Eq.~\eqref{eq:circ_ext} can be represented as a quantum channel $\Lambda_\tau$ acting on the ancilla:
\begin{equation}
\Qcircuit @C=1em @R=1em {
\lstick{\ket{\phi}} & \ctrl{1} &  \rstick{\rho_{f}}\qw \\
\lstick{\ket{\Psi}}   & \gate{U_\tau} &  \qw}\quad\quad\equiv\quad\quad
\Qcircuit @C=1em @R=1em {
\lstick{\rho_{i}} & \gate{\Lambda_\tau} &  \rstick{\rho_{f}}\qw}\quad\quad\;,
\end{equation}
where the output state of the ancilla is indicated here as a density matrix $\rho_f$. We want to show now how by performing quantum process tomography~\cite{Chuang1997,Altepeter2003,Mohseni2006,Mohseni2008} on the ancilla we can extract $\kappa_I$.

We can represent the quantum channel $\Lambda_\tau$ using the following Kraus decomposition (see eg. \cite{NandC_book})
\begin{equation}
\Lambda_\tau\left[\rho\right] = A_0 \rho A_0^\dagger + A_1 \rho A_1^\dagger
\end{equation}
with
\begin{equation}
A_0 = \begin{pmatrix}
1 & 0 \\
0 & \kappa \\
\end{pmatrix} \quad\quad A_1 = \begin{pmatrix}
0 & 0 \\
0 & \nu \\
\end{pmatrix}\;,
\end{equation}
but this choice is not unique. A better parametrization of the channels that overcomes this difficulty is to use the Pauli Transfer Matrix~\cite{Chow2012} defined as
\begin{equation}
R_{ij} = \frac{1}{2} Tr\left[P_i\Lambda_\tau\left[P_j\right]\right]
\end{equation}
where the $P_i$'s are the Pauli operators $\{\mathbb{1},X,Y,Z\}$. For our quantum channel this matrix takes the form
\begin{equation}
\label{eq:err_chan}
R_{U_\tau} = \begin{pmatrix}
1 & 0 &0 & 0\\
0 & \kappa_R &-\kappa_I&0\\
0 & \kappa_I &\kappa_R&0\\
0 & 0 &0 & 1\\
\end{pmatrix}
\end{equation}
where $\kappa = \kappa_R + i \kappa_I$. This form makes it apparent that the channel $\Lambda_\tau$ is a composition of a dephasing (or phase damping~\cite{NandC_book}) channel
\begin{equation}
R_z = \begin{pmatrix}
1 & 0 &0 & 0\\
0 & 1-p_z &0&0\\
0 & 0 &1-p_z&0\\
0 & 0 &0 & 1\\
\end{pmatrix}
\end{equation}
with error probability $p_z=\lvert\nu\rvert^2=1-\lvert\kappa\rvert^2$ and a rotation around the z-axis with angle $\theta=tan^{-1}\left(\kappa_I/\kappa_R\right)$ described by
\begin{equation}
R_\theta = \begin{pmatrix}
1 & 0 &0 & 0\\
0 & cos(\theta) &-sin(\theta)&0\\
0 & sin(\theta) &cos(\theta)&0\\
0 & 0 &0 & 1\\
\end{pmatrix}\;.
\end{equation}

As shown by Wiebe et al. in \cite{Wiebe2014} a Bayesian reconstruction strategy can be effectively employed to learn $\kappa_R$ and $\kappa_I$ from measurement of the device even in the presence of substantial depolarizing noise described by the channel
\begin{equation}
\Lambda_D(\rho) = (1-p_D) + \frac{p_D}{d}\mathbb{1}
\end{equation}
where $d=2^n$ and $n$ is the number of qubits used to encode $\ket{\Psi}$. According to the results in \cite{Wiebe2014}, large values $p_D=50\%$ could be handled with relative ease.

Furthermore, the reconstruction is simplified in our case since the structure of our channel is known beforehand, and one can tailor strategies aimed at estimating matrices of the form Eq.~\eqref{eq:err_chan}. The extent to which these could be leveraged to minimize the negative effect of more realistic noise channels in the system qubits is left for future explorations. Before concluding, we want to point out that the possibility of removing read-out lines from all but the ancilla qubit could also help more generally in reducing the overall noise in the device.

\section{Implementation Challanges}
\label{sec:imp_ch}
During our exposition of our methodology in Sec.~\ref{sec:newmethod} we have only briefely touched upon the practical cost of implementing the core parts of the algorithm on near term quantum devices. This section is dedicated to address those issues.
In particular we first present a discussion on how to estimate the the potential gain of using sQPE using condition Eq.~\eqref{eq:cond} in practical situations where only partial information on the high order coefficients $m_k$ is available. We then provide a description of the resources needed to implement the time evolution needed for sQPE (cf. Eq.~\eqref{eq:circuitcH}) using different strategies.

\subsection{Practical bound estimation}
\label{sec:pcond}
Due to the presence of the the expectation value $\langle O^{2K+1}\rangle$, it is difficult in most situations to asses directly if the condition in Eq.~\eqref{eq:cond} holds. In order to obtain a more manageable condition we can rewrite Eq.~\eqref{eq:cond} as
\begin{equation}
\label{eq:pconA}
\frac{\lvert\langle O\rangle\rvert}{\|\overline{O_T}\|_1} \geq \frac{f(K)^K}{\epsilon_r}\frac{\langle O^{2K}\rangle\lvert\langle O\rangle\rvert +Cov[O^{2K},O]}{\|\overline{O_T}\|^{2K+1}_1}\;.
\end{equation}
Due to fact that $h(x)=x^K$ with $K\geq1$ has a bounded first derivative on a finite interval $\Omega$ we have
\begin{equation}
\lvert x^{K} - y^{K}\rvert \leq \max_{z\in\Omega} [Kz^{K-1}] \lvert x - y\rvert\;,
\end{equation}
which in turn implies
\begin{equation}
 Var[O^{2K}]\leq 4K^2 {\lambda^{4K-2}_{max}} Var[O]
\end{equation}
with $\lambda_{max}$ the largest singular value of the operator $O$. Using the bound
\begin{equation}
\label{eq:cjensen}
\lvert Cov[X,Y] \rvert \leq \sqrt{Var[X]Var[Y]}\;,
\end{equation}
which can be obtained by using Jensen's inequality, we arrive at the following, looser, condition
\begin{equation}
\label{eq:pconB}
\frac{\lvert\langle O\rangle\rvert}{\|\overline{O_T}\|_1} \geq \frac{f(K)^K}{\epsilon_r}\frac{\langle O^{2K}\rangle\lvert\langle O\rangle\rvert +\|\overline{O}\|^{2K-1}_1Var[O]}{\|\overline{O_T}\|^{2K+1}_1}\;.
\end{equation}
Reasonably tight ubberbounds on $Var[O]$ and $\langle O^{2K} \rangle$ for small $K=\mathcal{O}(1)$ can be obtained in many situations of interest. An important example are many-body calculations of ground-state properties where a variational calculation with a classically simulatable trial states can provide such bounds with reasonable efficiency (eg. one could use Quantum Monte Carlo methods \cite{Foulkes2001,Carlson2015}). 
Another situation is when we have some control on the fidelity of the prepared state. For instance consider the case where we are preparing an initial state $\ket{\Psi}$ with large overlap with some eigenvector $\ket{\phi}$ of $O$
\begin{equation}
\rvert\Psi\rangle = \alpha \rvert \phi\rangle +\beta\rvert \phi^\perp\rangle\quad O\rvert\phi\rangle=\lambda_\phi\rvert\phi\rangle\;,
\end{equation}
with $\langle\phi\vert\phi^\perp\rangle=0$ and $\alpha^2+\beta^2=1$. We can obtain a bound on $\langle O^{2K} \rangle$ if we have a upperbound for the state fidelity
\begin{equation}
F\left[\ket{\Psi}\right]=Tr\left[\rvert\Psi\rangle\langle\Psi\vert\phi\rangle\langle\phi\lvert\right]=\lvert\langle\Psi\vert\phi\rangle\rvert^2<\Delta\;,
\end{equation}
by using
\begin{equation}
\langle O^{2K}\rangle \leq \lambda_\phi^{2K}+\Delta \|\overline{O}\|^{2K}\;.
\end{equation}

When only a bound on the variance is available instead we can use the inequality
\begin{equation}
 \langle O^{2K}\rangle\lvert\langle O\rangle\rvert\leq\lambda_{max}^{2K+1}\leq \|\overline{O}\|^{2K+1}_1
\end{equation}
to obtain the even looser condition
\begin{equation}
\label{eq:pconC}
\frac{\lvert\langle O\rangle\rvert}{\|\overline{O_T}\|_1} \geq \frac{f(K)^K}{\epsilon_r} \frac{\|\overline{O}\|^{2K+1}_1}{\|\overline{O_T}\|^{2K+1}_1}\left(1+\frac{Var[O]}{\|\overline{O}\|^2_1}\right)\;.
\end{equation}
Unfortunately this condition can bee too loose to be of practical value as we can see by looking at the limit $Var[O]\to0$: the approximate expression Eq.~\eqref{eq:pconB} recovers the correct limit Eq~\eqref{eq:cond_eig} as does the exact expression Eq.~\eqref{eq:pconA}, while the right hand side of Eq.~\eqref{eq:pconC} is always larger than $f(k)^K/\epsilon_r$ and this can produce an overly pessimistic assessment of the efficiency gain achievable with the sQPE scheme of Sec.~\ref{sec:newmethod}.

\begin{figure}
    \centering
    \includegraphics[scale=0.3]{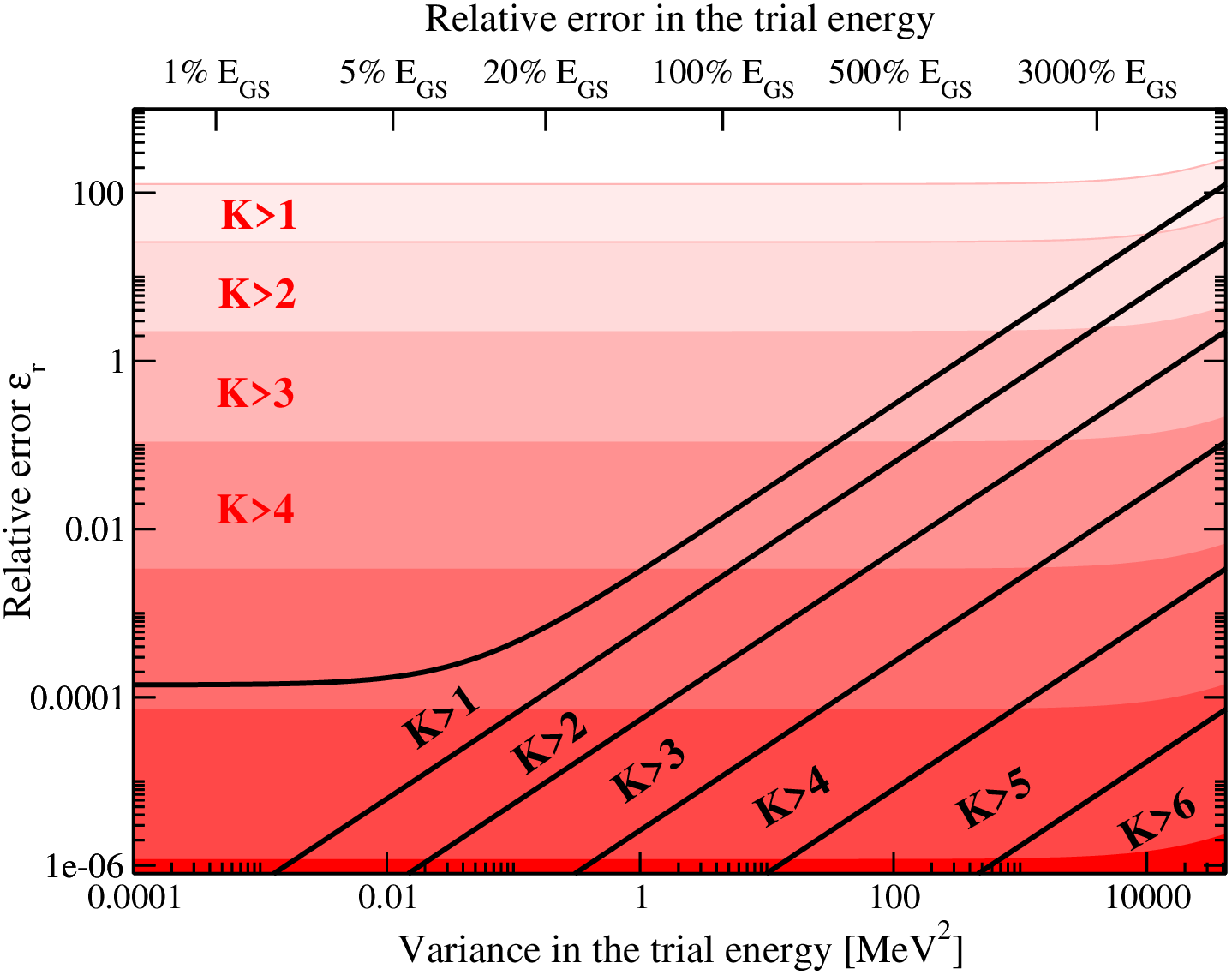}
    \caption{Estimated parameter regions for problems expected to be efficiently solvable with the scheme of Sec.~\ref{sec:newmethod} as predicted by Eq.~\eqref{eq:pconB} (black solid lines) and by the looser condition Eq.~\eqref{eq:pconC} (red contours). See text for more details.
    \label{fig:ubound}}
\end{figure}

To get more insight on this problem we plot in Fig.~\ref{fig:ubound} the regions in a two-dimensional $\left(Var[O],\epsilon_r\right)$ space where the inequalities above predict an advantage of our proposed scheme for the estimation of the deuteron's ground state energy. The solid black lines define the maximum relative error $\epsilon_r$ achievable for a trial state $\ket{\Psi}$ with a given variance obtained using the condition of Eq.~\eqref{eq:pconB}. The simple linear algorithm of Sec.~\ref{sec:lin} is predicted to be efficient for situations that stay in the top left corner of parameter space bounded by the first black line. For lower target $\epsilon_r$ or larger variance $Var[O]$ we then progressively need to increase the order $K$ of the algorithm to ensure Eq.~\eqref{eq:pconB} is satisfied. 
From Fig.~\ref{fig:ubound} we also see that for a $1\%$ target error the simple linear scheme with $K=1$ is predicted to be more efficient up to $Var[O]\approx E^2_{GS}$ before the growth of the bias term forces us to increase the order in $K$. This is encouraging since this condition is not necessarily tight in the sense that, due to the use of the upperbound Eq.~\eqref{eq:cjensen}, the inequality of Eq.~\eqref{eq:pconB} is a sufficient but not necessary condition for Eq.~\eqref{eq:pconA} to hold true. The extent to witch one can still find an efficiency gain by using the linear algorithm past this condition will most likely depend on the particular problem instance and is therefore difficult to predict without some prior knowledge on $Cov[O^{2K},O]$ that would allow to use Eq.~\eqref{eq:pconA}.

The red contours in Fig.~\ref{fig:ubound} are obtained instead by using the looser condition Eq.~\eqref{eq:pconC} and as we can see they are overly pessimistic: a relatively high order algorithm with $K>4$ is judged to be needed even in the limiting case when the trial state has zero variance.
This observation reinforces the importance of being able to use a tighter condition like Eq.~\eqref{eq:pconA} in order to assess meaningfully the possibility of a gain in using the strategy proposed in this work.

\subsection{Time evolution}
\label{sec:time_evo}
We finally turn our attention to the problem of estimating the circuit depth required for the implementation of the time-evolution unitary $U_\tau=e^{i\tau O}$ needed for the sQPE method. In particular we consider the more realistic situation where we only have an approximation $\widetilde{U_\tau}$ of $U_\tau$ with error bounded by $\delta_\tau$ is available:
\begin{equation}
\| \widetilde{U_\tau}-U_\tau\|\leq \delta_\tau\;.
\end{equation}
In this situation the induced error on the sQPE estimator Eq.~\eqref{EKtau} is then $\epsilon_\tau=\delta_\tau/\tau$. A simple way to control the total error $\epsilon$ of the calculation is to require that $\epsilon_\tau<\epsilon/2$ and similarly for the MSE $\epsilon_M$ in Eq.~\eqref{eq:mse}. The latter modification will increase the bound reported in Eq.~\eqref{eq:NT} only by a factor $2^{1+1/K}$. This requirement can be relaxed if one employs optimal algorithms like \cite{Berry2015b,Low16,Low17} that allow to implement $\widetilde{U_\tau}$ for $\epsilon_\tau\ll \epsilon$ with only a small increase in gate count.

Due to the relatively short propagation time required by the sQPE algorithm, we can also use simpler strategies based on the Trotter--Suzuki decomposition~\cite{Suzuki91} while still maintaining a short circuit depth. To see how this works let us start by considering a simple first order scheme obtained by dividing the propagation time $\tau$ into $r$ segments:
\begin{equation}
T(\tau,r) = e^{i\tau\alpha_0} \left[\prod_{k=1}^L e^{i\frac{\tau}{r} \alpha_k U_k}\right]^r
\end{equation}
where we used the decomposition Eq.~\eqref{eq:lcu} for the operator $O$. As shown in~\cite{Childs18} the error for this approximation can be bounded by
\begin{equation}
\label{eq:prop_ebound}
\left\| T(\tau,r)-U_\tau\right\|\leq \frac{\left(\tau\|\overline{O}\|_1\right)^2}{r}e^{\frac{\tau}{r}\|\overline{O}\|_1}
\end{equation}
with the the norm $\|\overline{O}\|_1=\lvert\alpha_0\rvert+\|\overline{O_T}\|_1$ defined as in Eq.~\eqref{eq:qnorm}. Note that this is slightly tighter than the result Proposition F.3 obtained in~\cite{Childs18} and can be found following the same proof. We can now provide the following analytic bound for the number of intervals $r$ required to guarantee that $\epsilon_\tau<\epsilon/2$
\begin{equation}
\label{eq:lin_interval}
r_1 = \left\lceil \max{\left[\tau\|\overline{O}\|_1,\frac{2e}{\epsilon\tau}\left(\tau\|\overline{O}\|_1\right)^2\right]} \right\rceil
\end{equation}
where we used the same derivation as in~\cite{Childs18}. For convenience we rewrite the optimal time step Eq.~\eqref{eq:time_step} as
\begin{equation}
\tau_{opt} =\gamma(K)\left(\frac{\epsilon}{\rvert m_K\rvert} \right)^{{\frac{1}{2K}}}\;,
\end{equation}
where we defined
\begin{equation}
\gamma(K)\equiv\left(\frac{(2K+1)!}{2\sqrt{2K+1}}\right)^{\frac{1}{2K}}\;.
\end{equation}
Note the additional factor of $2$ in the denominator coming from the choice $\epsilon_M=\epsilon/2$. We can now express the bound on $r$ for the simple linear product formula as
\begin{equation}
\label{eq:r1bound}
r_1 = \left\lceil \rho_1 \max{\left[ 1, \frac{2e}{\epsilon}\|\overline{O}\|_1\right]} \right\rceil
\end{equation}
where we have defined
\begin{equation}
\rho_1=\tau_{opt}\|\overline{O}\|_1 = \gamma(K)\frac{\|\overline{O}\|_1}{\lvert m_K\rvert^{\frac{1}{2K}}}\epsilon^{\frac{1}{2K}}\;,
\end{equation}
and for reasonably small errors the bound Eq.~\eqref{eq:r1bound} is maximized with the rightmost expression giving the algorithm an overall depth scaling at best as $r_1=\mathcal{O}\left(1/\sqrt{\epsilon}\right)$.
It is instructive to express these bounds in terms of the relative error $\epsilon_r$ and the expectation value ratio $R_O$ from Eq.~\eqref{eq:ratio}, for instance in the case of eigenvalue estimation we find
\begin{equation}
\rho_1=\gamma(K)\frac{\|\overline{O}\|_1}{\|\overline{O_T}\|_1} \frac{\epsilon_r^{\frac{1}{2K}}}{R_O}
\end{equation}
which leads to a generic scaling given by
\begin{equation}
r_1= \mathcal{O}\left(\gamma(K)\frac{\|\overline{O}\|^2_1}{\|\overline{O_T}\|^2_1} \frac{\epsilon_r^{\frac{1-2K}{2K}}}{R_O^2}\right)
\end{equation}
which in the linear case of Sec.~\ref{sec:lin} simplifies to
\begin{equation}
r_1= \mathcal{O}\left(\frac{\|\overline{O}\|^2_1}{\|\overline{O_T}\|^2_1} \frac{1}{R_O^2}\frac{1}{\sqrt{\epsilon_r}}\right)\;.
\end{equation}
This provides only a minor advantage over the $\mathcal{O}\left(1/\epsilon\right)$ scaling associated with full fledged QPE algorithms~\cite{Knill07} which can be easily spoiled with a sufficiently small $R_O$ ratio. It is therefore important to use higher order expansions that are able to achieve a more favourable scaling and in the following we will consider higher order product formulas as an example.
If we denote the ($2j$)-th order Trotter-Suzuki formula with $r$ intervals~\cite{Suzuki91,Childs18} as $S_{2j}(\tau,r)$ we can generalize the error bound Eq.~\eqref{eq:prop_ebound} obtained above to
\begin{equation}
\label{eq:hprop_ebound}
\left\| S_{2j}(\tau,r)-U_\tau\right\|\leq \frac{\left(2\tau5^{j-1}\|\overline{O}\|_1\right)^{2j+1}}{3r^{2j}}e^{2\frac{\tau}{r}5^{j-1}\|\overline{O}\|_1}
\end{equation}
and bound the number of intervals as
\begin{equation}
r_{j} = \left\lceil\rho_{j} \max{\left[1,\left(\frac{4e}{3\epsilon}5^{j-1}\|\overline{O}\|_1\right)^{\frac{1}{2j}}\right]} \right\rceil
\end{equation}
with $\rho_{j}\equiv2\rho_1 5^{j-1}$. Again this is slightly tighter than the result obtained by Childs et al. in the Supplemental Material of~\cite{Childs18}. As for the linear decomposition the right term dominates for reasonably small errors $\epsilon$ and we find the overall scaling
\begin{equation}
r_{j} = \mathcal{O}\left(5^{j+\frac{1}{2j}} \frac{\gamma(K)}{\lvert m_K\rvert^{\frac{1}{2K}}}\|\overline{O}\|_1^{1+\frac{1}{2j}} \epsilon^{\frac{j-K}{4jK}}\right)\;.
\end{equation}
%In the more common case where the optimal time step is not known and approximated instead using the upperbound %$\Gamma_K$ from Eq.~\eqref{eq:gamma_ratio} we find
%\begin{equation}
%r_{2K} = \mathcal{O}\left(5^{j+\frac{1}{2j}} \frac{\gamma(K)}{\Gamma_K^{\frac{1}{2K}}}\|\overline{O}\|_1 %\left(\frac{\epsilon}{\|\overline{O}\|_1}\right)^{\frac{j-K}{4jK}}\right)\;.
%\end{equation}
As above we can express this in terms of relative quantities in a compact way for the special case of eigenvalue estimation
\begin{equation}
r_{j} = \mathcal{O}\left( 5^{j+\frac{1}{2j}}\gamma(K) \left(\frac{\|\overline{O}\|_1}{\|\overline{O_T}\|_1}\right)^{1+\frac{1}{2j}} \frac{\epsilon_r^{\frac{j-K}{4jK}}}{R_O^{1+\frac{1}{2j}}}\right)\;,
\end{equation}
and due to the fast growth of the first term in the above expression we might want to keep the order $j$ as small as possible. For instance using $j=K$ will already guarantee a gate count independent on the target precision $\epsilon$ and scaling as $\mathcal{O}\left(1/R_O^{3/2}\right)$ in terms of the eigenvalue ratio. Notably by simply choosing $j=K+1$, for the price of a fixed increase in cost of less than a factor of $5$ we can achieve a circuit depth that decreases as a function of the target relative error.

Note that these estimates are based on the possibly very pessimistic bounds Eq.~\eqref{eq:prop_ebound} and Eq.~\eqref{eq:hprop_ebound} which means that these circuit depths could possibly be greatly reduced in practice (see eg.~\cite{Childs18}). Before finishing this section we want to point out that even though the estimates provided above are for the implementation of $U_\tau$ a complete implementation of it's controlled version needed in Eq.~\eqref{eq:circuitcH} can be obtained with an overall linear increase in depth as a function of the number of qubits in the system register used to represent $\ket{\Psi}$. Tighter bounds will require further knowledge of the particular operator $O$ whose time evolution we want to simulate, but the results presented here give us good reasons to believe practical implementations could be achievable for interesting systems on near term devices.

\section{Summary and Conclusions}
\label{sec:conc}

In this work we reviewed the standard methodology of Operator Averaging~\cite{Peruzzo2014,McClean2014,McClean2016} to evaluate expectation values $\langle O\rangle$ of general Hermitian operators $O$ efficiently on quantum computers by minimizing the number of quantum operations needed while maintaining a shot-noise limited number of measurement $N_{tot}=\mathcal{O}\left(1/\epsilon^2\right)$ to attain precision $\epsilon$ in the expectation value estimate. This provides a great advantage on current generation noisy devices where the asymptotically optimal behaviour $N_{tot}=\mathcal{O}\left(1/\epsilon\right)$ of methods that employ Quantum Phase Estimation~\cite{Abrams1999,Knill07} cannot be attained in practice due to the large circuit depths $C_{QPE}=\mathcal{O}\left(1/\epsilon\right)$ involved.
As we explain in Sec.~\ref{sec:SM} however, the Operator Average strategy has a major drawback in that in terms of relative error $\epsilon_r$ the total measurement count grows as $N_{tot}=\mathcal{O}\left(1/(\epsilon_r R_O)^2\right)$ where $R_O$ defined in Eq.~\eqref{eq:ratio} is approximately the ratio between the wanted expectation value and the largest eigenvalue of $O$. 

In this work we propose to use a single step of phase estimation as in the well known Hadamard Test to learn the expectation value $\langle O\rangle$ by looking at the short time behaviour of $\langle sin(\tau O)\rangle$ instead. This strategy was already discussed in the context of full QPE calculations in~\cite{Knill07} and it remains the method of choice for fully error-corrected devices capable of executing accurately very long gate sequences. Our contribution is in showing how, by using circuits implementing only a single Hadamard Test with appropriately chosen time-steps, one can greatly reduce the classical cost (the number of experimental measurements $N_{tot}$) using much shorter circuits than those needed for QPE. For instance in the important case of eigenvalue estimation we can achieve $N_{tot}=\mathcal{O}(1/\epsilon_r^{2+1/K})$ for $K=\mathcal{O}(1)$ independent on $R_O$ while keeping the gate count bounded by $C_{sQPE}\mathcal{O}\left(5^K/R_O^{1+1/2K}\right)$ using a very simple general purpose strategy employing the high-order Trotter-Suzuki decomposition~\cite{Suzuki91,Childs18}. As we argue in Sec.~\ref{sec:time_evo} the latter requirement can possibly be further reduced by using more advanced simulation strategies~\cite{Berry2015b,Low16,Low17} and we plan to further this possibility along the lines of the study presented in~\cite{Childs18} in a future work.

We presented a complete analysis of the first two lowest order sQPE algorithms with $K=1$ and $K=2$ in Sec.~\ref{sec:newmethod} together with a self consistent procedure aimed at finding the optimal time-steps to be used in the calculation. As our approach could be extremely helpful in some situations but it is not efficient in an asymptotic scaling sense in general, we provide both strict and easy to estimate conditions to help determine if the use of sQPE con provide a speedup for a particular problem instance (see Sec.~\ref{sec:newmethod} and Sec.~\ref{sec:pcond}). As these conditions require the availability of bounds on the expectation value to be computed and some control over the operator spectrum (like bounds on the $n$-th cumulant $\langle O^n\rangle$) further work on classically efficient strategies to estimate them (using for instance ideas from~\cite{Baumgratz2012,Harrow2017}) could have a possible big impact on the practicality of our approach. As discussed in Sec.~\ref{sec:pcond} classical Quantum Monte Carlo simulations could be employed efficiently in the meantime.
Finally in Sec.~\ref{sec:noise} we have shown some evidence on the robustness of our proposed methodology to readout noise on the quantum device and provided arguments to justify the expectation that ancilla based algorithms like sQPE provide in general a much more robust layout to deploy and execute non trivial quantum algorithms on NISQ devices. It will be very interesting in the future to see the impact of adaptive machine-learning techniques as those presented in~\cite{Wiebe2014} on the practical feasibility of scaling up quantum computations in the near term.

\begin{acknowledgments}
We wish to thank N. Klco and R. Schiavilla for helpful discussions regarding the subject of this work. We also thank A. Matsuura of Intel Labs for useful conversations and for encouragements. The work of A.R. was supported by the U.S. Department of Energy, Office of Science, Office of Advanced Scientific Computing Research (ASCR) quantum algorithm teams and testbed programs, under field work proposal number ERKJ333 and by U.S. Department of Energy grant No. DE-FG02-00ER41132. We also grateful to the T2 group at Los Alamos National Laboratory for the hospitality while working on this project thanks in part to funds from the U.S. Department of Energy, Office of Science, HEP Contract No.DE-KA2401032. A.B. acknowledges support from U.S. Department of Energy, Office of Science, Office of Nuclear Physics, under Award No. DE-SC0010300 and DE-SC0019647.
\end{acknowledgments}

\bibliography{biblio}   

\appendix
\section{Implementation of controlled time-evolution}
\label{app:circuit}
We report in this section the implementation of the two-qubit controlled time evolution appearing in Eq.~\eqref{eq:circuitcH} needed for the sQPE algorithm and schematically presented in Eq.~\eqref{eq:circ_ext} in the main text. Throughout this section we assume that the initial state $\ket{\Psi}$ has been prepared with a rotation $R_y(\theta)$ with $\theta$ the angle of interest.
For a system of two qubits a generic controlled unitary operation associated with a $2\times2$ unitary matrix $U$ can be represented in the computational basis $\rvert 00\rangle$, $\rvert 01\rangle$,$\rvert 10\rangle$, $\rvert 11\rangle$ as
\begin{equation}
\label{eq:cu}
C_U=\begin{pmatrix}
1_{2\times2}&0_{2\times2}\\
0_{2\times2}& U\\
\end{pmatrix}\; ,
\end{equation}
with $1_{2\times 2}$ and $0_{2\times 2}$ indicating the two-by-two identity matrix null matrix respectively.

Let's now recall the general decomposition of a $U(2)$ unitary
\begin{equation}
U = e^{i\theta_0}R_z(\theta_1)R_y(\theta_2)R_z(\theta_3)
\end{equation}
for appropriately chosen angles. The rotation matrices here are defined as
\begin{align}
R_z(\phi)&=\begin{pmatrix}
e^{i\phi/2} & 0 \\
0 & e^{-i\phi/2} \\
\end{pmatrix} \\
R_y(\theta)&=\begin{pmatrix}
\cos(\theta/2) & \sin(\theta/2) \\
-\sin(\theta/2) & \cos(\theta/2) \\
\end{pmatrix}\;.
\end{align}

Using this decomposition, together with the definition Eq.~\eqref{eq:cu}, we can implement the controlled time-evolution operator with the following circuit (see eg.~\cite{Barenco95})
\begin{widetext}
\begin{equation}
\Qcircuit @C=1em @R=.7em {
&\qw&\gate{\Phi(\theta_0)}&\qw&\ctrl{1}&\qw&\qw&\ctrl{1} &\qw&\qw\\
&\qw&\gate{R_z(\theta_1)} &\gate{R_y\left(\frac{\theta_2}{2}\right)}&\targ &\gate{R_y\left(-\frac{\theta_2}{2}\right)}& \gate{R_z(-\frac{\theta_3+\theta_1}{2})}& \targ &\gate{R_z(\frac{\theta_3-\theta_1}{2})}&\qw\\
	}
\end{equation}
%or even better with the following one
%\begin{equation}
%\Qcircuit @C=1em @R=.7em {
%&\qw&\gate{\Phi(\theta_0)}&\ctrl{1}&\qw&\qw&\ctrl{1} &\qw&\qw&\qw\\
%&\qw&\gate{R_z\left(\frac{\theta_1-\theta_3}{2}\right)} &\targ %&\gate{R_z\left(-\frac{\theta_1+\theta_3}{2}\right)}& \gate{R_y\left(-\frac{\theta_2}{2}\right)}& \targ %&\gate{R_y\left(\frac{\theta_2}{2}\right)}&\gate{R_z\left(\theta_3\right)}&\qw\\
%	}
%\end{equation}
\end{widetext}
%The latter one is advantageous because we can drop the last two rotations.
where we have defined the phase gate as
\begin{equation}
\Phi(\theta)=\begin{pmatrix}
1&0\\
0&e^{i\theta}
\end{pmatrix}\;.
\end{equation}
Using the following decomposition of the Hamiltonian matrix
\begin{equation}
H=\begin{pmatrix}
\alpha&\beta\\
\beta&\gamma\\
\end{pmatrix} \equiv \frac{\alpha+\gamma}{2}\mathbb{1} + \beta X+ \frac{\alpha-\gamma}{2}Z\label{eq:ham_gen}
\end{equation}
with $X,Z$ Pauli spin matrices, $\mathbb{1}$ the identity matrix and $(\alpha,\beta,\gamma)$ real numbers, we can write the exact time-propagator as
\begin{equation}
e^{i\delta H}=e^{i\delta\frac{\alpha+\gamma}{2}}\left[cos(\theta)+i\hat{\theta}\cdot\sigma sin(\theta)\right]
\label{eq:exp}
\end{equation}
with
\begin{equation}
\vec{\theta}=(\delta\beta,0,\delta\frac{\alpha-\gamma}{2})\quad\hat{\theta}=\frac{\vec{\theta}}{\theta}\;.
\end{equation}
From this expression we can easily determine the needed angles $(\theta_0,\dots,\theta_3)$.
The full circuit for sQPE is then
\begin{widetext}
\begin{equation}
\Qcircuit @C=1em @R=.7em {
\lstick{\ket{0}}&\qw&\gate{H}&\gate{\Phi(\theta_0)}&\qw&\ctrl{1}&\qw&\qw&\ctrl{1} &\qw&\gate{S}&\gate{H}&\qw&\meter\\
\lstick{\ket{\Psi}}&\qw&\qw&\gate{R_z(\theta_1)} &\gate{R_y\left(\frac{\theta_2}{2}\right)}&\targ &\gate{R_y\left(-\frac{\theta_2}{2}\right)}& \gate{R_z(-\frac{\theta_3+\theta_1}{2})}& \targ &\gate{R_z(\frac{\theta_3-\theta_1}{2})}&\qw&\qw&\qw&\qw\\
	}
\end{equation}
\end{widetext}
and the last rotation can be avoided.
\end{document}